\documentclass[fleqn,usenatbib]{mnras}

\usepackage{newtxtext,newtxmath}
\usepackage[T1]{fontenc}
\usepackage{ae,aecompl}
\usepackage{graphicx}
\graphicspath{{Figures/}}
\usepackage{tabularx}
\usepackage{amsmath}
\usepackage{amssymb}
\usepackage{wasysym}

\newcolumntype{C}{>{\centering\arraybackslash}X}
\newcolumntype{L}{>{\raggedright\arraybackslash}X}
\newcolumntype{R}{>{\raggedleft\arraybackslash}X}

\title[A cool core census]{A census of cool core galaxy clusters in IllustrisTNG}

\author[D. J. Barnes et al.]{David J. Barnes$^{1,2}$\thanks{E-mail: djbarnes@mit.edu},
Mark Vogelsberger$^{1}$\thanks{Alfred P. Sloan Fellow},
Rahul Kannan$^{1,3}$\thanks{Einstein Fellow},
Federico Marinacci$^{1}$,
\newauthor
Rainer Weinberger$^{4}$,
Volker Springel$^{4,5,6}$,
Paul Torrey$^{1}$\thanks{Hubble Fellow},
Annalisa Pillepich$^{7}$,
\newauthor
Dylan Nelson$^{6}$,
R\"udiger Pakmor$^{4}$,
Jill Naiman$^{3}$,
Lars Hernquist$^{3}$,
Michael McDonald$^{1}$
\\
$^1${Department of Physics, Kavli Institute for Astrophysics and Space Research, Massachusetts Institute of Technology, Cambridge, MA 02139, USA}\\
$^2$Jodrell Bank Centre for Astrophysics, School of Physics and Astronomy, The University of Manchester, Manchester M13 9PL, UK\\
$^3${Harvard--Smithsonian Center for Astrophysics, 60 Garden Street, Cambridge, MA 02138}\\
$^4${Heidelberg Institute for Theoretical Studies, Schloss-Wolfsbrunnenweg 35, D-69118 Heidelberg, Germany}\\
$^5${Zentrum f{\"u}r Astronomie der Universit{\"a}t Heidelberg, ARI, M{\"o}nchhofstr. 12-14, D-69120 Heidelberg, Germany}\\
$^6${Max-Planck-Institut f{\"u}r Astrophysik, Karl-Schwarzschild-Str. 1, 85741 Garching, Germany}\\
$^7${Max-Planck-Institut f{\"u}r Astronomie, K{\"o}nigstuhl 17, 69117 Heidelberg, Germany}\\
}

\date{Accepted XXX. Received YYY; in original form ZZZ}

\pubyear{2017}

\begin{document}
\label{firstpage}
\pagerange{\pageref{firstpage}--\pageref{lastpage}}
\maketitle

\begin{abstract}
The thermodynamic structure of hot gas in galaxy clusters is sensitive to astrophysical processes and typically difficult to model with galaxy formation simulations. We explore the fraction of cool-core (CC) clusters in a large sample of $370$ clusters from IllustrisTNG, examining six common CC definitions. IllustrisTNG produces continuous CC criteria distributions, the extremes of which are classified as CC and non-cool-core (NCC), and the criteria are increasingly correlated for more massive clusters. At $z=0$, the CC fractions for $2$ criteria are in reasonable agreement with the observed fractions but the other $4$ CC fractions are lower than observed. This result is partly driven by systematic differences between the simulated and observed gas fraction profiles. The simulated CC fractions with redshift show tentative agreement with the observed fractions, but linear fits demonstrate that the simulated evolution is steeper than observed. The conversion of CCs to NCCs appears to begin later and act more rapidly in the simulations. Examining the fraction of CCs and NCCs defined as relaxed we find no evidence that CCs are more relaxed, suggesting that mergers are not solely responsible for disrupting CCs. A comparison of the median thermodynamic profiles defined by different CC criteria shows that the extent to which they evolve in the cluster core is dependent on the CC criteria. We conclude that the thermodynamic structure of galaxy clusters in IllustrisTNG shares many similarities with observations, but achieving better agreement most likely requires modifications of the underlying galaxy formation model.
\end{abstract}


\section{Introduction}
\label{sec:intro}
Galaxy clusters are the most massive collapsed structures at the current epoch; forming hierarchically under gravity via the accretion of matter and merging with other collapsed haloes. This formation process shock-heats the intracluster medium (ICM) to $10^{7}-10^{8}\,\mathrm{K}$, resulting in the emission of X-rays. Early X-ray observations of the ICM revealed that the cooling time in the cores of some clusters was significantly shorter than the Hubble time \citep{Lea1973,FabianNulsen1977,Cowie1977,Mathews1978}. Clusters with short central cooling times are also associated with more relaxed morphologies and drops in central temperatures, reaching only a third of the virial temperature \citep{Ikebe1997,Lewis2002,Peterson2003,Vikhlinin2005}. These systems are known as cool-core (CC) clusters \citep{Molendi2001}.

The fraction of the cluster population that host a CC depends strongly on the method of sample selection, due to the so-called `cool-core bias' \citep{Eckert2011}. CCs are associated with a strong peak in their X-ray surface brightness and a higher X-ray luminosity at fixed mass compared to non-cool-core (NCC) clusters, making them more easily detected in flux-limited X-ray samples. Hence, the fraction of CCs in X-ray samples is likely overestimated. For example, $\approx60$ per cent of $207$ clusters detected with \textit{Einstein}, over $59$ per cent of $55$ clusters observed with \textit{ROSAT} and $72$ per cent of $64$ clusters in the HIFLUGCS sample were found to host a CC \citep{White1997,Peres1998,Hudson2010}. However, the success of Sunyaev-Zel'dovich (SZ) surveys \citep{Hasselfield2013,Bleem2015,Planck2016} has enabled the collection of mass-limited samples that avoid cool-core bias \citep[see][]{Lin2015}. Using the Planck SZ sample, \citet{Planck_ER_XI} found that $35$ per cent of the sample hosted a CC, \citet{Rossetti2017} found a CC fraction of $29$ per cent and \citet{Andrade-Santos2017} found that, depending on the defining criterion, $28-39$ per cent of Planck clusters hosted a CC.

Observational results suggest that the properties of CCs have evolved little since $z\sim1.6$. The central electron number density, entropy and cooling time of CC clusters at low-redshift are similar to their high-redshift counterparts \citep{Cavagnolo2009,McDonald2013,McDonald2017}, suggesting that thermal equilibrium in the cluster core is established very early in its formation history. These results are consistent with the observation that the fraction of strong CCs, defined by cuspiness of the electron number density \citep{Vikhlinin2007} or concentration of the surface brightness \citep{Santos2008,Santos2010}, decreases with increasing redshift. This is because the majority of the cluster volume evolves self-similarly around a CC, increasing in density and reducing the contrast to the core with increasing redshift \citep{McDonald2017}.

Reproducing the observed CC/NCC fractions in numerical simulations has been a significant challenge \citep{BorganiKravtsov2011}. Idealized simulations have examined the physical processes in idealized CC setups \citep[e.g.][]{McCourt2012,Sharma2012,Gaspari2015,Li2015}, but have not reproduced how CCs form or how they are maintained in a cosmological setting. Cosmological simulations give conflicting results on CC formation and maintenance. \citet{Kay2007} found that almost all clusters hosted a CC at both $z=0$ and $z=1$. \citet{Burns2008} also found that the fraction of CCs as a function of redshift was roughly constant, but that only $\approx15$ per cent of systems hosted a CC. The predicted fraction of CC clusters is inconsistent with the observed fraction and the lack of evolution in the fraction of CC systems is in tension with the mild evolution that is observed \citep[e.g.][]{McDonald2017}. Early work also proposed that the difference between CCs and NCCs was driven by activity at high redshift $(z>1)$, with significant early mergers \citep{Burns2008} or preheating \citep{McCarthy2008} resulting in the production of NCC clusters and late time mergers being unable to turn a CC into a NCC \citep{Poole2008}. This is in contrast to more recent numerical work that has shown that late time mergers are capable of destroying CCs \citep{Rasia2015}, with \citet{Hahn2017} advocating that the angular momentum of the merger is critical in determining whether the CC is disrupted. \citet{Rasia2015} argue that the lack of AGN feedback in early work, which leads to overcooling, made the cores too resilient to late time mergers. In addition, the conflicting numerical results are further complicated by the use of different criteria when defining a CC cluster.

In this paper, we examine the fraction of CC and NCC galaxy clusters in the IllustrisTNG simulations, a follow-up to the Illustris project that contains an updated galaxy formation model and larger simulation volumes. Compared to previous theoretical work our study has the advantages of: being a significantly larger sample of clusters than many previous studies; being at higher numerical resolution compared to many previous works; employing a state-of-the-art galaxy formation model that shows good agreement with observational results from dwarf galaxies to cluster scales. This enables us to study the fraction of clusters defined as CC and NCC for a variety of criteria commonly used in the literature. We examine the fraction of CC clusters produced by the IllustrisTNG model at $z=0$ and compare the criteria distributions to those from low-redshift observational samples \citep{Cavagnolo2009,Andrade-Santos2017}. We then study how the CC fraction evolves with redshift for different criteria and compare the fraction of CCs and NCCs defined as relaxed as a function of redshift. Finally, we examine how cluster cores evolve compared to the rest of the cluster volume by comparing the hot gas profiles at $z=0$ and $z=1$.

The paper is structured as follows. In Section \ref{sec:sim} we briefly describe the IllustrisTNG model and the simulation volume we use. We present the CC criteria that we use throughout this work in Section \ref{sec:CCcrit}. In Section \ref{sec:results} we examine the criteria as a function of mass and the correlation between them. We then investigate how the CC fraction, defined by the different criteria, evolves with redshift and the fraction of CC systems that are defined as relaxed in Section \ref{sec:CCevo}. In Section \ref{sec:profs} we study how the profiles of CCs evolve compared to NCCs. We then present our conclusions in Section \ref{sec:concs}.

\section{Numerical Method}
\label{sec:sim}
IllustrisTNG \citep{Springel2017,Marinacci2017,Naiman2017,Pillepich2017b,Nelson2017} is a follow-up project to the Illustris simulation \citep{Vogelsberger2014,VogelsNat2014,Genel2014,Sijacki2015}, which reproduces the galaxy size-mass relation \citep{Genel2017} and the metallicity content of the ICM \citep{Vogelsberger2017}. The IllustrisTNG suite contains three simulation volumes, TNG50, TNG100 and TNG300, each at three different resolution levels (1, 2 and 3). All simulations use a cosmological model with parameters chosen in accordance with the \citet{PlanckXIII2016} constraints: $\Omega_{\mathrm{m}}=0.3089$, $\Omega_{\mathrm{b}}=0.0486$, $\Omega_{\Lambda}=0.6911$, $\sigma_{8}=0.8159$, $H_{0}=100\,h\,\mathrm{km}\,\mathrm{s}^{-1}\mathrm{Mpc}^{-1}=67.74\mathrm{km}\,\mathrm{s}^{-1}\mathrm{Mpc}^{-1}$ and $n_{\mathrm{s}}=0.9667$.

In this work, we analyze the clusters that are present in the TNG300-1 periodic volume. The TNG300-1 volume has a side length of $302.6\,\mathrm{Mpc}$ and a dark matter and baryonic mass resolution of $5.9\times10^{7}\,\mathrm{M}_{\astrosun}$ and $1.1\times10^{7}\,\mathrm{M}_{\astrosun}$ respectively. The collisionless particles, i.e. stars and dark matter, have a softening length of $1.48\,\mathrm{kpc}$, which is comoving for $z>1$, and a fixed physical length for $z\leq1$. The gas cells employ an adaptive comoving softening length that reaches a minimum of $0.37\,\mathrm{kpc}$. The simulation was performed with the moving-mesh code \textsc{Arepo} \citep{Springel2010}, and evolved the magneto-hydrodynamics equations \citep{Pakmor2013}. IllutrisTNG employs an updated version of the Illustris galaxy formation model \citep{Vogelsberger2013,Torrey2014,Vogelsberger2014,VogelsNat2014,Genel2014}, a comprehensive set of subgrid physical models that now include a new radio mode AGN feedback scheme \citep{Weinberger2017}, a re-calibrated SN wind model and an extension to the chemical evolution scheme  \citep{Pillepich2017} and refinements to the numerical scheme to improve the convergence properties \citep{Pakmor2016}.

\renewcommand\arraystretch{1.2}
\begin{table*}
  \caption{Table summarizing the criteria used throughout this work with the CC and MCC definitions.}
 \begin{tabularx}{13.9cm}{l c l l l}
 \hline
 Criterion & Notation & Aperture & CC limit & MCC limit \\
 \hline
 Central electron number density & $n_{\mathrm{e}}$ & $0.012\,r_{500}$ & $>1.5\times10^{-2}\,\mathrm{cm}^{-3}$ & $>0.5\times10^{-2}\,\mathrm{cm}^{-3}$ \\
 Central cooling time & $t_{\mathrm{cool}}$ & $0.012\,r_{500}$ & $<1\,\mathrm{Gyr}$ & $<7.7\,\mathrm{Gyr}$ \\
 Central entropy excess & $K_{0}$ & -- & $<30\,\mathrm{keV}\,\mathrm{cm}^{-2}$ & $<60\,\mathrm{keV}\,\mathrm{cm}^{-2}$ \\
 Concentration parameter (physical) & $C_{\mathrm{phys}}$ & $40.0,~400.0\,\mathrm{kpc}$ & $>0.155$ & $>0.075$ \\
 Concentration parameter (scaled) & $C_{\mathrm{scal}}$ & $0.15,~1.0\,r_{500}$ & $>0.5$ & $>0.2$ \\
 Cuspiness parameter & $\alpha$ & $0.04\,r_{500}$ & $>0.75$ & $>0.5$ \\
 \hline
 \end{tabularx}
 \label{tab:CCbounds}
\end{table*}
\renewcommand\arraystretch{1.0}

Haloes were identified via a standard friends-of-friends (FoF) algorithm with a linking length of $b=0.2$ and substructures were identified using \textsc{subfind} \citep{Springel2001,Dolag2009}. We analyze a mass-limited sample and select all haloes with $M_{500}>10^{13.75}\,\mathrm{M}_{\astrosun}$\footnote{We define $M_{500}$ as the mass enclosed within a sphere of radius $r_{500}$ whose mean density is $500$ times the critical density of the Universe at the cluster's redshift.} at each redshift. At $z=0$ $(z=1)$ this yields a sample of $370$ $(77)$ clusters with a median mass of $M_{500}=8.8\times10^{13}\,\mathrm{M}_{\astrosun}$ $(M_{500}=8.2\times10^{13}\,\mathrm{M}_{\astrosun})$. X-ray luminosities, $L_{\mathrm{X}}$, are computed using the Astrophysical Plasma Emission Code \citep[\textsc{apec};][]{Smith2001} via the \textsc{pyatomdb} module with atomic data from \textsc{atomdb} v3.0.3 \citep[last described in][]{Foster2012}. We compute mock X-ray spectra for each individual chemical element tracked by the simulation in the rest-frame energy band $0.05-100\,\mathrm{keV}$ for each gas cell, using its density, temperature, and metallicity. We then sum the spectra for the elements to produce an X-ray spectrum for each cell. X-ray luminosity within an aperture is then calculated by summing the spectra in the desired energy band for the gas cells that fall within the aperture, see \citet{LeBrun2014,Barnes2017a} for further details. The luminosity within a 2D aperture is computed by collapsing the 3D positions of the gas cells along the $z$-axis and then summing the luminosities of all gas cells within the FoF group whose 2D position is within the aperture. We have explored the impact of collapsing the positions along different axes or using the median or mean value from the three different axes and find that it makes negligible difference to the CC criteria distributions presented throughout this work.

\section{Cool-core criteria}
\label{sec:CCcrit}
The literature contains many ways of defining a CC cluster and the fraction of clusters defined as CC depends on the chosen criterion and the sample selection. Observationally, the choice of criterion used will depend on the quality and resolution of the data, for example, the ability to extract a temperature profile. In this section we define the CC criteria that we will consider in this work.\footnote{The thresholds for different CC criteria are taken from the references within each section.} The thresholds for defining a cluster as CC or moderate cool core (MCC) for the different criteria are summarized in Table \ref{tab:CCbounds}. When calculating the criteria we only include non-star forming gas that is cooling (i.e. not being heated via supernovae or AGN feedback) and has a temperature $T>1.0\times10^{6}\,\mathrm{K}$. 

The choice of halo centre can impact the recovered CC fraction and in this work we always centre on the cluster's potential minimum. \citet{McDonald2013} demonstrated that for strong CC centering via the gas centroid or X-ray peak yields the same result. In addition, the observational sample of \citet{Cavagnolo2009} and \citet{Andrade-Santos2017} only centre using the X-ray peak. Therefore, for consistency we always select X-ray peak centred data when we have the choice. 

\subsection{Central electron number density}
A rapidly cooling core will be colder than its surrounding material and, therefore, must be denser than its surroundings in order to maintain pressure equilibrium. Observationally, the electron number density is extracted from the measured surface brightness profile and requires significantly fewer counts than other CC criteria, such as the cooling time. Following \citet{Hudson2010}, we define a cluster as CC if the average density $n_{\mathrm{e}}(<0.012r_{500}) > 1.5\times10^{-2}\,\mathrm{cm}^{-3}$. We choose to measure the quantity inside a 3D aperture of $0.012r_{500}$ to mimic observational apertures \citep[e.g.][]{Hudson2010,McDonald2017,Andrade-Santos2017}. Those clusters with a central density in the range $0.5\times10^{-2}<n_{\mathrm{e}}(<0.012r_{500})\leq1.5\times10^{-2}\,\mathrm{cm}^{-3}$ are defined as MCC clusters and the remaining are defined as NCC.

\subsection{Central cooling time}
Classically, clusters have been defined as CCs if the cooling time of their central region is short compared to the age of the Universe. Cooling time is defined as
\begin{equation}\label{eq:cooling}
 t_{\mathrm{cool}}=\frac{3}{2}\frac{(n_{\mathrm{e}}+n_{\mathrm{i}})k_{\mathrm{B}}T}{n_{\mathrm{e}}n_{\mathrm{i}}\Lambda(T,Z)}\,,
\end{equation}
where $ n_{\mathrm{i}}$ is the ion number density, $k_{\mathrm{B}}$ is Boltzmann's constant, $T$ is the temperature of the gas and $\Lambda$ is the cooling function, which is a function of temperature and metallicity, $Z$. Mimicking observations, we measure the average cooling time inside a 3D aperture of $r<0.012r_{500}$ \citep[e.g.][]{McDonald2013}. We define those clusters with a cooling time less than $1\,\mathrm{Gyr}$ as CCs. A common definition of a NCC cluster is that its central cooling time is more than half the age of the Universe, hence at $z=0$ $t_{\mathrm{cool}}\approx7.7\,\mathrm{Gyr}$. However, at $z=1$ half the age of the Universe implies a cooling time $t_{\mathrm{cool}}\approx2.9\,\mathrm{Gyr}$. Therefore, following \citet{McDonald2013} we set the threshold for clusters to be considered as MCCs to $t_{\mathrm{cool}}<7.7\,\mathrm{Gyr}$ at all redshifts to make the definition independent of redshift. \citet{McDonald2013} tested the impact of the different definitions and found it to negligible. Those clusters with $t_{\mathrm{cool}}\geq7.7\,\mathrm{Gyr}$ are defined as NCCs.

\subsection{Central entropy excess}\label{eq:entropy}
The central entropy excess of a cluster is another method of defining whether it is a CC or not. To measure the central entropy excess we follow the method of \citet{Cavagnolo2009}. We first compute the 3D radial entropy profile using $50$ radial bins in the range $10^{-3}-1.5\,r_{500}$, where the entropy $K=k_{\mathrm{B}}Tn_{\mathrm{e}}^{-2/3}$. We then fit the entropy profile in the range $0.01-1.0r_{500}$ with a power-law of the form
\begin{equation}
 K(r) = K_{0}+K_{100}\left(\frac{r}{100\,\mathrm{kpc}}\right)^{a}\,,
\end{equation}
where $K_{0}$ is the excess entropy above the best-fitting power-law, $K_{100}$ is the normalization of the entropy at $100\,\mathrm{kpc}$, $a$ is the power-law index and $r$ is the radial distance from the cluster's potential minimum. We then use $K_{0}$ as the measure of the central entropy excess. We define CC clusters as those with $K_{0}<30\,\mathrm{keV}\,\mathrm{cm}^{2}$, MCCs as those in the range $30\leq K_{0}<60\,\mathrm{keV}\,\mathrm{cm}^{2}$ and the remaining as NCCs.

\subsection{X-ray concentration parameter}
As the X-ray emissivity is proportional to the gas density squared, and only weakly depends on temperature, CC clusters should have X-ray bright cores. Therefore, the ratio of the X-ray luminosity within the core compared to the luminosity within a larger aperture should clearly demonstrate the presence of a CC. First proposed by \citet{Santos2008}, this criterion is commonly known as the concentration parameter. We define the concentration parameter as
\begin{equation}\label{eq:Cphys}
 C_{\mathrm{phys}} = \frac{L_{\mathrm{X}}^{\mathrm{soft}}(r_{\mathrm{p}}<40\,\mathrm{kpc})}{L_{\mathrm{X}}^{\mathrm{soft}}(r_{\mathrm{p}}<400\,\mathrm{kpc})}\,,
\end{equation}
where $L_{\mathrm{X}}^{\mathrm{soft}}$ is the soft band X-ray luminosity in the energy range $0.5-5.0\,\mathrm{keV}$ and $r_{\mathrm{p}}$ is the projected radial distance from the cluster's potential minimum. Those clusters with $C_{\mathrm{phys}}>0.155$ are defined as CCs, those in the range $0.075<C_{\mathrm{phys}}\leq0.155$ as MCCs, and the rest as NCCs. We also consider the modification to this parameter proposed by \citet{Maughan2012}, a scaled concentration parameter where the apertures are scaled by the cluster's characteristic radius $r_{500}$,
\begin{equation}\label{eq:Cfrac}
 C_{\mathrm{scal}} = \frac{L_{\mathrm{X}}^{\mathrm{soft}}(r_{\mathrm{p}}<0.15r_{500})}{L_{\mathrm{X}}^{\mathrm{soft}}(r_{\mathrm{p}}<r_{500})}\,.
\end{equation}
For the scaled concentration parameter we define CCs as those with $C_{\mathrm{scal}}>0.5$, MCCs as $0.2<C_{\mathrm{scal}}\leq0.5$, and the rest as NCCs.

\subsection{Cuspiness parameter}
First proposed by \citet{Vikhlinin2007}, the cuspiness of the electron number density profile is a common metric for defining CC clusters. Following the literature, we calculate the cuspiness parameter of the 3D density profile, computed using $50$ radial bins in the range $10^{-3}-1.5\,r_{500}$, via
\begin{equation}\label{eq:alpha}
 \alpha = -\left.\frac{d\log n_{\mathrm{e}}(r)}{d\log r}\right|_{r=0.04r_{500}}\,.
\end{equation}
We define CCs as those with $\alpha>0.75$, MCCs in the range $0.5<\alpha\leq0.75$ and the remaining are classified as NCCs. We have used the CC threshold from \citet{Hudson2010} that is marginally higher than the $\alpha>0.7$ used in \citet{Vikhlinin2007}, but this choice makes negligible difference to the results presented in this work.

Using these common CC criteria we now examine the fraction of CC clusters produced by the IllustrisTNG model at $z=0$.

\section{Low-redshift cool-cores}
\label{sec:results}
The limited volume of the TNG300 simulation means that the overlap in the mass distributions of the observed and simulated samples is limited, with the observed clusters being more massive on average than the simulated clusters. Therefore, here we split the sample in three mass bins: low-mass $(M_{500}<9\times10^{13}\,\mathrm{M}_{\astrosun})$, intermediate-mass $(9\times10^{13}\leq M_{500}<2.0\times10^{14}\,\mathrm{M}_{\astrosun})$ and high-mass ($M_{500}\geq2\times10^{14}\,\mathrm{M}_{\astrosun})$, and compare the complete sample and the high-mass sample to the observed CC fraction. At $z=0$, the mass bins contain $191$, $139$ and $49$ clusters and have median $M_{500}$ values of $6.9\times10^{13}\,\mathrm{M}_{\astrosun}$, $1.2\times10^{14}\,\mathrm{M}_{\astrosun}$ and $2.7\times10^{14}\,\mathrm{M}_{\astrosun}$, respectively. To explore the dependence of the CC criteria on the total mass of the halo we fit a simple linear relation of the form
\begin{equation}\label{eq:linear}
 y = mx+c\:,
\end{equation}
and we measure the scatter about the best-fit for each mass bin via
\begin{equation}\label{eq:scatter}
 \sigma_{\log_{10}} = \sqrt{\frac{1}{N}\sum_{i=1}^{N}\left[\log_{10}(y_{i})-\log_{10}(y_{\mathrm{mod}})\right]^{2}}\,,
\end{equation}
where $N$ is the number of clusters in the sample, $Y_{i}$ is the measured criterion value, $Y_{\mathrm{mod}}$ is the expected criterion value for a cluster with a given $M_{500}$ and we note that $\sigma_{\ln}=\ln(10)\sigma_{\log_{10}}$. The fit and scatter uncertainties are computed by bootstrapping the sample $10,000$ times.

\subsection{CC fractions}
\begin{figure*}
 \includegraphics[width=1.025\textwidth,keepaspectratio=True]{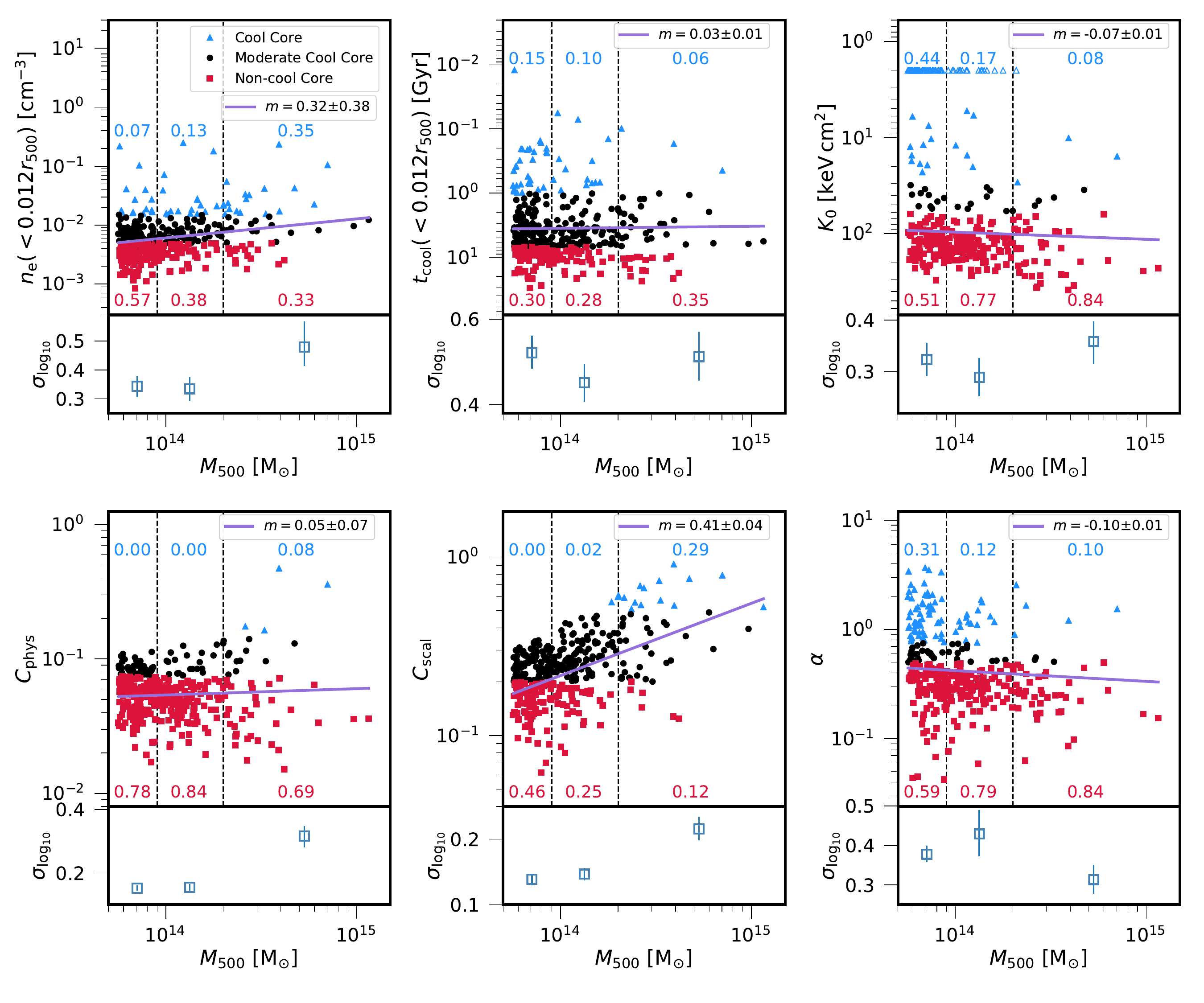}
 \caption{Comparison of the six CC criteria as a function of $M_{500}$ at $z=0$ (upper panel) and the scatter relative to the best-fit relation for the three mass bins (lower panel). We plot the central electron number density (top left), central cooling time (top centre), central entropy excess (top right), the concentration parameter within physical apertures (bottom left) and scaled apertures (bottom centre), and the cuspiness parameter (bottom right). CCs are denoted by blue triangles, MCCs by black circles and NCCs by red squares. We set $K_{0}=2\,\mathrm{keV}\,\mathrm{cm}^{2}$ for those clusters whose central entropy excess falls below this value, making them visible on the plot, and denote these systems by open symbols. The solid purple line denotes the best-fit linear relation. The dashed black lines denote the mass bins and the (red) blue numbers give the fraction of (non-)cool-core systems in each bin. We note the $y$-axis is inverted for the central cooling time and central entropy excess, such that CCs always appear at the top of a panel.}
 \label{fig:M500-CCcrit}
\end{figure*}

\renewcommand\arraystretch{1.1}
\begin{table*}
 \caption{Table of the fraction of clusters defined as CC, MCC or NCC for the criteria presented in Section \ref{sec:CCcrit} for the complete sample and the different mass bins at $z=0$. All errors are computed by bootstrap resampling 10,000 times.}
 \begin{tabularx}{\textwidth}{l C C C C C C C}
 \hline
 Sample & Fraction & \multicolumn{6}{c}{Criteria} \\
 $(z=0)$ & & $n_{\mathrm{e}}$ & $t_{\mathrm{cool}}$ & $K_{0}$ & $C_{\mathrm{phys}}$ & $C_{\mathrm{scal}}$ & $\alpha$ \\
 \hline
 Complete & CC & $0.14\pm0.02$ & $0.12\pm0.02$ & $0.18\pm0.02$ & $0.01\pm0.01$ & $0.04\pm0.01$ & $0.21\pm0.02$ \\
 $(M_{500}>10^{13.75}\,\mathrm{M}_{\astrosun})$ & MCC & $0.39\pm0.03$ & $0.58\pm0.03$ & $0.10\pm0.01$ & $0.20\pm0.02$ & $0.62\pm0.02$ & $0.09\pm0.01$ \\
 $N=370$ & NCC & $0.46\pm0.02$ & $0.30\pm0.03$ & $0.72\pm0.02$ & $0.79\pm0.02$ & $0.34\pm0.03$ & $0.70\pm0.02$ \\
 \hline
 Low-mass & CC & $0.07\pm0.02$ & $0.15\pm0.03$ & $0.28\pm0.04$ & $0.00\pm0.01$ & $0.00\pm0.01$ & $0.31\pm0.04$ \\
 $(M_{500}<9.0\times10^{13}\,\mathrm{M}_{\astrosun})$ & MCC & $0.36\pm0.03$ & $0.55\pm0.03$ & $0.12\pm0.02$ & $0.22\pm0.04$ & $0.54\pm0.04$ & $0.10\pm0.03$ \\
 $N=191$ & NCC & $0.57\pm0.04$ & $0.30\pm0.04$ & $0.60\pm0.04$ & $0.78\pm0.03$ & $0.46\pm0.04$ & $0.59\pm0.04$ \\
 \hline
 Intermediate-mass & CC & $0.13\pm0.03$ & $0.10\pm0.02$ & $0.05\pm0.03$ & $0.00\pm0.01$ & $0.02\pm0.02$ & $0.12\pm0.03$ \\
 $(9.0\times10^{13}\leq M_{500}<2.0\times10^{14}\,\mathrm{M}_{\astrosun})$ & MCC & $0.49\pm0.05$ & $0.62\pm0.05$ & $0.06\pm0.02$ & $0.16\pm0.04$ & $0.73\pm0.04$ & $0.09\pm0.03$ \\
 $N=130$ & NCC & $0.38\pm0.05$ & $0.28\pm0.05$ & $0.88\pm0.02$ & $0.84\pm0.03$ & $0.25\pm0.04$ & $0.79\pm0.03$ \\
 \hline
 High-mass & CC & $0.35\pm0.06$ & $0.06\pm0.04$ & $0.10\pm0.06$ & $0.08\pm0.06$ & $0.29\pm0.06$ & $0.10\pm0.06$ \\
 $(M_{500}\geq2.0\times10^{14}\,\mathrm{M}_{\astrosun})$ & MCC & $0.33\pm0.06$ & $0.59\pm0.06$ & $0.14\pm0.06$ & $0.22\pm0.06$ & $0.59\pm0.06$ & $0.06\pm0.05$ \\
 $N=49$ & NCC & $0.33\pm0.06$ & $0.35\pm0.06$ & $0.76\pm0.05$ & $0.69\pm0.06$ & $0.12\pm0.06$ & $0.84\pm0.07$ \\
 \hline
 \end{tabularx}
 \label{tab:CCfracs}
\end{table*}
\renewcommand\arraystretch{1.0}

In Fig. \ref{fig:M500-CCcrit} we plot the different CC criteria presented in Section \ref{sec:CCcrit} as a function of $M_{500}$ at $z=0$. Clusters that are defined as CC are denoted by blue triangles, MCCs as black circles and NCCs as red squares. The CC metrics all produce continuous distributions with no obvious bimodality or dichotomy, suggesting that CC and NCC clusters are the two extremes of the same distribution. In Appendix \ref{app:resolution} we examine the impact of numerical resolution on the criteria distributions using the level 1 and level 2 resolutions of the TNG300 simulation, finding that numerical resolution has minimal impact on the criteria presented here.

In the upper left panel of Fig. \ref{fig:M500-CCcrit}, we plot the central electron number density as a function of $M_{500}$. The complete sample of clusters yields a CC fraction of $14\pm2$ per cent, a MCC fraction of $39\pm2$ per cent and a NCC fraction of $46\pm2$ per cent. For the high-mass sample we find an increased CC fraction of the $35\pm6$ per cent. We compare to the observed central electron number densities of the clusters present in the \textit{Planck} Early Sunyaev-Zel'dovich (ESZ) sample \citep{Andrade-Santos2017}. Selecting clusters with $z<0.25$, the observed sample has a CC fraction of $41\pm4$ per cent. Therefore, the CC fraction recovered for the high-mass sample is in reasonable agreement with the observed CC fraction. Fitting the criterion distribution with the linear relation given in eq. \ref{eq:linear} we find a slope $m=0.32\pm0.38$, i.e. consistent with no mass dependence. However, measuring the scatter within each mass bin we find that the scatter increases significantly for the high-mass bin.

The central cooling time as a function of $M_{500}$ is plotted in the upper centre panel of Fig. \ref{fig:M500-CCcrit}. Defining CCs via this criterion we find that $12\pm2$ per cent of clusters are defined as CC for the complete sample. In contrast to the central electron number density, the high-mass bin has a CC fraction of $6\pm4$ per cent, a factor $2$ lower than the complete sample. We compare to the observed clusters that form the \textit{ACCEPT} sample \citep{Cavagnolo2009} with $z<0.25$, noting that the median mass of the observed sample is more massive than the simulated sample. The observed clusters yield a CC fraction of $52\pm4$ per cent. However, the \textit{ACCEPT} clusters are taken from the \textit{Chandra} archive and in effect form an X-ray selected sample, that suffers from CC bias. \citet{Andrade-Santos2017} quantify the selection effects by assuming that the mass function and X-ray luminosity-mass relation are power laws. Using values from \citet{Vikhlinin2009a} and \citet{Vikhlinin2009b} they estimate that CCs are over-represented by a factor of $\sim2.2$. This reduces the observed CC fraction to $0.24$ per cent, which is a factor $4$ larger than CC fraction produced by the high-mass sample. Fitting the simulated criterion distribution as a function of mass we find a very weak mass dependence with a slope of $m=0.03\pm0.01$ and the scatter across the three mass bins is consistent within the errors.

CCs are defined by the central entropy excess in the upper right panel of Fig. \ref{fig:M500-CCcrit}. This yields a CC fraction of $18\pm2$ per cent for the complete sample. Many of the simulated clusters defined as CC have very small central entropy excesses, which increases the simulated CC fraction and is further discussed below. To make these clusters visible on the plot we have set their central entropy excess to $2\,\mathrm{keV}\,\mathrm{cm}^{2}$ and denoted them by open triangles. The central entropy excess produces a decreasing CC fraction with increasing halo mass with the high-mass bin yielding a CC fraction of $10\pm6$ per cent, a factor $2$ smaller than the observed CC fraction. For the same CC definition clusters in the \textit{ACCEPT} sample with $z<0.25$ yield a CC fraction of $48\pm3$ per cent. However, we note that the limited resolution of the temperature profiles for some \textit{ACCEPT} clusters have been shown to induce an artificial floor in the entropy profile \citep[e.g.][]{Panagoulia2014,Hogan2017}. This leads to a second peak in the entropy excess distribution at $15-20\,\mathrm{keV}$. However, the true central entropy excess of these clusters would be lower than this artificial floor and these clusters would still be defined as CCs if the floor was removed. In addition, this second peak is $50-100$ per cent lower than our adopted CC threshold and we still classify these objects as CCs. However, the observed CC fraction should be viewed as a lower limit. Therefore, despite well documented issues, we compare to \textit{ACCEPT} sample as it provides a large-statistic observational dataset. Assuming that CCs are over-represented in this X-ray selected observational sample by a factor $2.2$ the observed CC fraction of $22$ per cent, a factor $2$ larger than the high-mass sample. However, we note that both the simulations and observations have significant shortcomings and this may impact the CC fractions presented here. We fit the simulated criterion distribution, excluding those clusters with $K_{\mathrm{0}}<2\,\mathrm{keV}\,\mathrm{cm}^{2}$, and find a weak mass dependence with a slope of $m=-0.07\pm0.01$ and a scatter that is consistent within the error across the mass bins.

In the left and central bottom panels of Fig. \ref{fig:M500-CCcrit} we plot the concentration parameter within physical, $C_{\mathrm{phys}}$, and scaled, $C_{\mathrm{scal}}$, apertures as function of $M_{500}$, respectively. Measured within physical apertures the concentration parameter produces a CC fraction of $1\pm1$ per cent for the complete sample, while within scaled apertures the complete sample yields a CC fraction of $4\pm1$ per cent. The high-mass sample yields CC fractions of $8\pm6$ per cent and $29\pm6$ per cent for the physical and scaled concentration parameter respectively. The clusters in the \textit{Planck} ESZ sample with $z<0.25$ produce CC fractions $36\pm5$ per cent and $28\pm4$ per cent for the physical and scaled concentration parameter, respectively. The high-mass bin is a factor $4.5$ lower than observed for the physical concentration parameter and consistent with the observations for the scaled concentration parameter. Fitting the simulated criteria distributions we find that the physical concentration parameter is consistent with no mass dependence $m=0.05\pm0.07$, while the scaled concentration parameter yields a slope of $m=0.41\pm0.04$ and increases with increasing halo mass. The scatter for both criteria increases significantly in the high-mass bin relative to the other bins.

\begin{figure*}
 \includegraphics[width=1.025\textwidth]{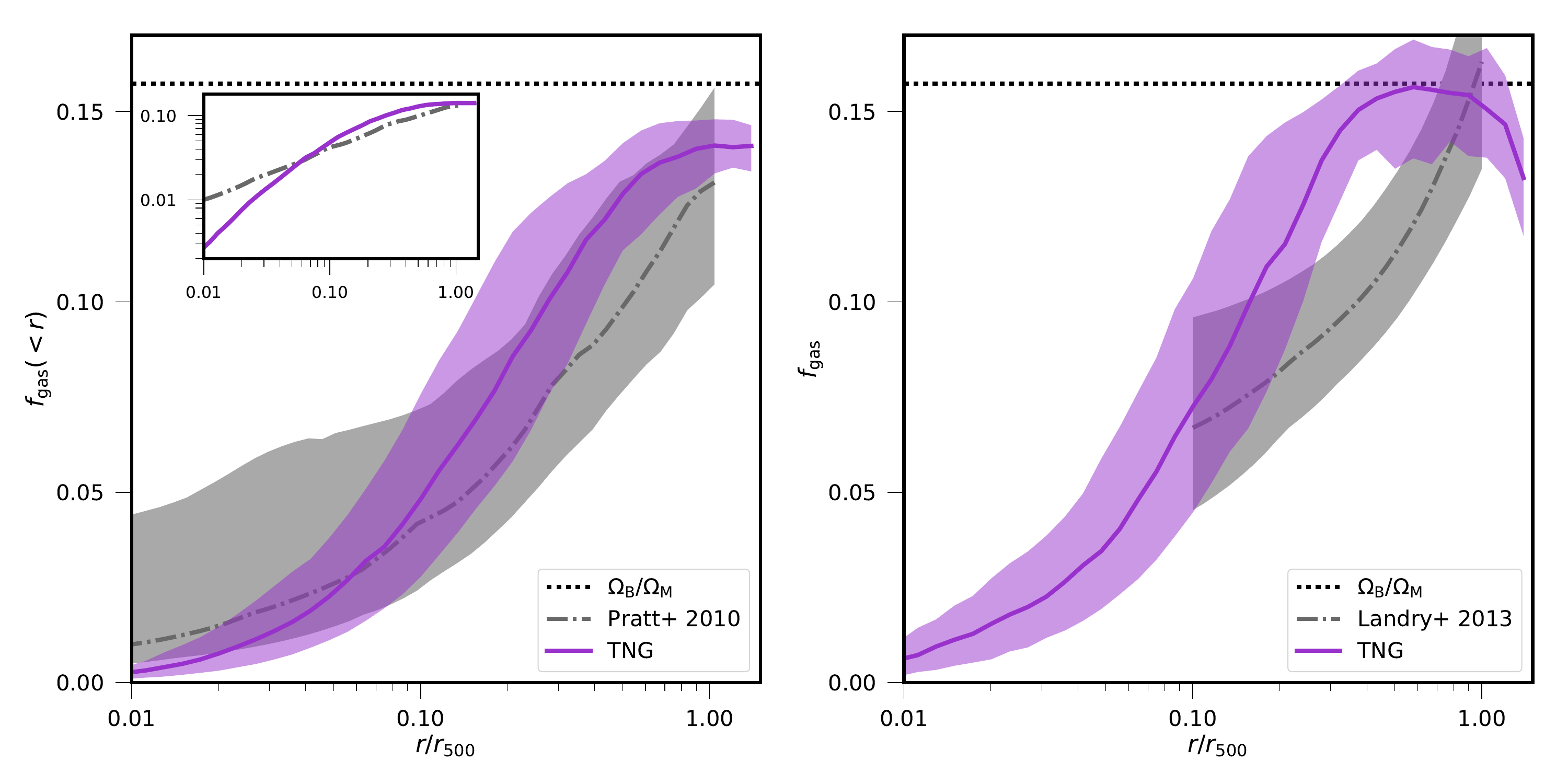}
 \caption{Median cumulative (left panel) and differential (right panel) gas fraction profiles as a function of $r/r_{500}$ at $z=0$. We plot the observed profiles from \citet{Pratt2010} and \citet{Landry2013} (dashed gray line) for the cumulative and differential profiles respectively, with the shaded regions encompassing $68$ per cent of the sample. To ensure a fair comparison, we compute the median simulated profile for those clusters with $M_{500}>2\times10^{14}\,\mathrm{M}_{\astrosun}$ (solid purple line). The universal baryon fraction $f_{\mathrm{b}}=(\Omega_{\mathrm{B}}/\Omega_{\mathrm{M}}\equiv0.157)$ is denoted by the black dotted line. The inset in the left panel shows the gas fraction profiles on a log scale to clarify the difference at small radii. The median simulated profiles rise more steeply than the observed profiles, before flattening at larger radii.}
 \label{fig:gas_fraction_comp}
\end{figure*}

Finally, in the bottom right panel of Fig. \ref{fig:M500-CCcrit} we plot the cuspiness parameter, $\alpha$, as a function of $M_{500}$. The complete sample yields a CC fraction of $21\pm2$ per cent, which is a factor $2$ larger than the CC fraction recovered for the high-mass bin. The clusters with $z<0.25$ in the \textit{Planck} ESZ sample yield a CC fraction of $35\pm5$ per cent, a factor $3.5$ larger than the high-mass sample. Fitting the simulated criterion distribution as a function of mass we find a mild negative mass dependence $m=-0.10\pm0.01$ with increasing halo mass and the scatter in the high-mass bin decreases relative to the other two mass bins. 

Compared to other recent numerical work with modern hydrodynamic solvers and more developed subgrid prescriptions, we find that IllustrisTNG yields similar CC fractions. \citet{Rasia2015} found at $z=0$ that $38$ per cent ($11/29$) of clusters were classified as CC, defined by pseudo-entropy and a central entropy excess criterion of $K_{0}<60\,\mathrm{keV}\,\mathrm{cm}^{2}$. Although we do not calculate a pseudo-entropy for our clusters, if we make the same cut based on the central entropy excess we find that $28\pm5$ per cent of the complete sample at $z=0$ are classified as CCs. In comparison, $57\pm4$ per cent of clusters with $z<0.25$ in the \textit{ACCEPT} sample are defined as CC, corrected to $26$ per cent assuming a CC bias correction factor of $2.2$. \citet{Hahn2017} defined a cluster as CC if the central entropy excess measured at $40\,\mathrm{kpc}$ was $K_{0}<40\,\mathrm{keV}\,\mathrm{cm}^{2}$. They find that $40$ per cent $(4/10)$ of their clusters at low-redshift $(z\leq0.37)$ are classified as CC. Making the same cut in the central entropy excess the complete sample produces a CC fraction of $19\pm2$ per cent at $z=0$ and $49\pm3$ per cent at $z=0.4$. With the same criterion the \textit{ACCEPT} cluster sample at $z<0.25$ yields a CC fraction of $50\pm3$ per cent ($23$ per cent corrected). The \textsc{c-eagle} project \citep{Barnes2017b,Bahe2017} resimulated $30$ clusters using the state-of-the-art \textsc{eagle} galaxy formation model \citep{Schaye2015,Crain2015}, at a similar resolution to the TNG100 level-1 volume. However, the clusters had low-density, high-entropy cores compared to the REXCESS sample, making it unlikely that any of these clusters would be classified as CCs. The low-density, high-entropy cores were thought to be due to the AGN feedback being ineffective at high redshift, resulting in some overcooling, and too active at late times, increasing the central entropy of the clusters. In summary, we find that the IllustrisTNG model yields a sample of clusters that, in general, has a lower CC fraction than observed when compared to low-redshift SZ selected samples, where the impact of observational selection effects are expected to be less than $1$ per cent \citep{Pipino2010,McDonald2013,Lin2015}. However, the simulated CC fractions are consistent with other recent numerical work.

\subsection{Cumulative gas fractions}
\label{ssec:fgas}
To further understand the lower fraction of CCs present in the IllustrisTNG volume we now examine the cumulative and differential gas fractions. In Fig. \ref{fig:gas_fraction_comp} we plot the median cumulative (left panel) and differential (right panel) gas fraction as a function of radius at $z=0$. We compare to the observed profiles from \citet{Pratt2010} and \citet{Landry2013} for the cumulative and differential profiles respectively. These have median masses of $M_{500}=2.8\times10^{14}\,\mathrm{M}_{\astrosun}$ and $M_{500}=5.4\times10^{14}\,\mathrm{M}_{\astrosun}$. We calculate the median simulated profile for those clusters with $M_{500}>2\times10^{14}\,\mathrm{M}_{\astrosun}$, which yields a sample with a median mass of $M_{500}=2.78\times10^{14}\,\mathrm{M}_{\astrosun}$. We note the observational sample is X-ray selected and includes several very strong CCs.

\begin{figure*}
 \includegraphics[width=\textwidth]{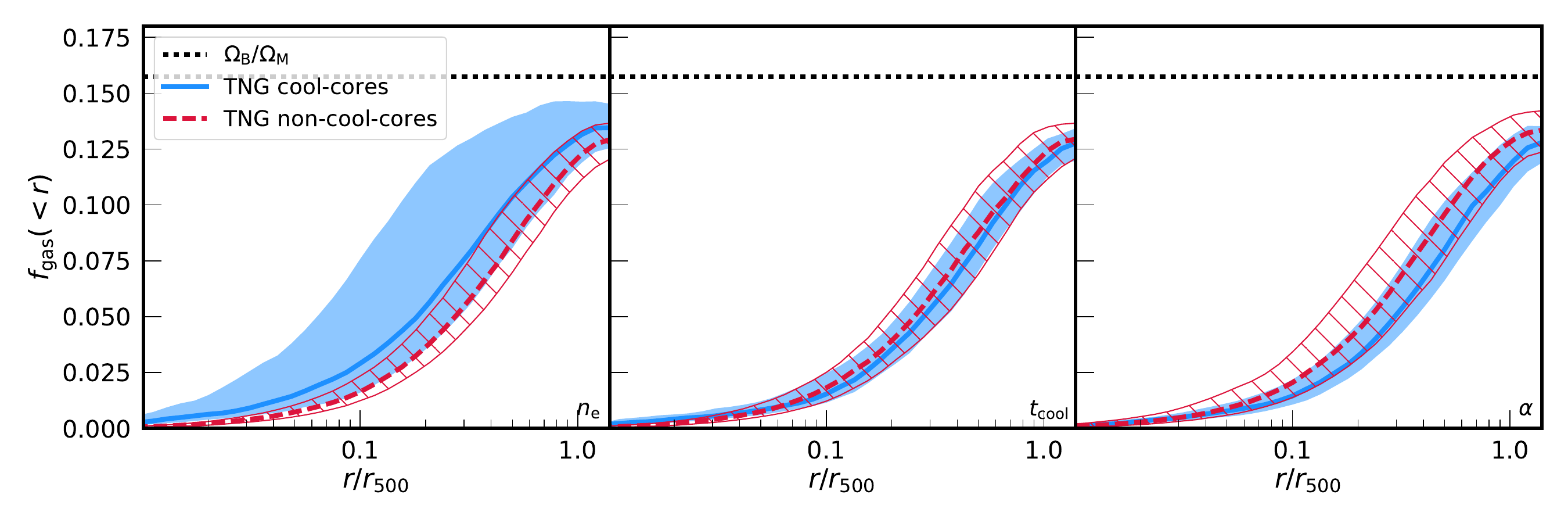}
 \caption{Median cumulative gas fractions at $z=0$ for CCs (solid blue line) and NCCs (red dashed line) defined by the central electron number density (left panel), central cooling time (centre panel) and cuspiness parameter (right panel). The shaded (hashed) region denotes $68$ per cent of the sample for CCs (NCCs) and the dotted line denotes the universal baryon fraction. We find that CCs and NCCs have very similar gas profiles regardless of the defining criterion.}
 \label{fig:fgas_CC_NCC}
\end{figure*}

The observed cumulative median gas fraction increases by a factor of $3.1$, from $28$ per cent of the universal baryon fraction $(\Omega_{\mathrm{B}}/\Omega_{\mathrm{M}}\equiv0.157)$ at $0.1r_{500}$ to $83$ percent at $r_{500}$. Although the median simulated profile increases by a similar factor between $0.1-1r_{500}$, it rises more rapidly and reaches $86$ per cent of the universal fraction at $0.6r_{500}$. Between $0.6-1r_{500}$ the median profile flattens to a constant value of $90$ per cent of the universal fraction. The simulated and observed differential profiles are similarly different, with the simulated profile rising much more rapidly than the observed profile and peaking at smaller radii compared to the observed profile. These differences between the observed and simulated profiles suggest that the simulated AGN feedback is more violent than in reality. This drives gas from the centre and results in a steepening of the gas fraction profiles, though we note that the feedback is significantly gentler than the previous Illustris model which removed the gas from the potential entirely \citep{Genel2014}. Additionally, the cumulative profiles show that the gas fraction within $r_{500}$ is higher than observed by $\sim10$ per cent. This may indicate that the AGN is not efficient enough at removing gas from the potential at high-redshift, and ejected gas then is reaccreted at low redshift. Similar gas fraction results were found in the \textsc{c-eagle} cluster simulations \citep{Barnes2017b}.

The gas fraction profiles help to explain why we find a lower simulated CC fraction than observed, especially for the concentration parameter. A gas fraction that rises more rapidly will result in a greater fraction of the X-ray emission coming from larger radii. This will result in systematically lower concentration parameter values. In the inset of Fig. \ref{fig:gas_fraction_comp} we plot the gas fraction on a log scale and we find that the median simulated gas mass enclosed in $0.01r_{500}$ is $70$ per cent lower than observed. Therefore, the central electron number density in the simulated clusters is systematically lower and fewer clusters will be defined as CCs by this criterion. From eq. \ref{eq:cooling}, it is clear that a lower central number density will also produce longer cooling times and result in fewer clusters being defined as CCs by the cooling time criterion.

We examine the difference between the cumulative gas mass profiles for CCs and NCCs for the complete sample in Fig. \ref{fig:fgas_CC_NCC}, where CCs are defined by the central electron number density (left panel), cooling time (middle panel) and cuspiness parameter (right panel) criteria. In general, we find that CC and NCC clusters have very similar profiles, regardless of defining criteria. This is in contrast to previous numerical work by \citet{Hahn2017} who found that their simulated CCs had significantly higher central gas fractions compared to NCCs, but in agreement with \citet{Eckert2013} who observed little difference between the CC and NCC gas fraction profiles of $62$ clusters. Defining CCs by their central cooling time, the median profiles and the region denoting $68$ per cent of the population are almost identical at all radii. Defining CCs by their central electron number density results in the CC profile having a higher median gas fraction at a fixed radius compared to the NCC median profile, $70$ per cent at $0.1r_{500}$, and larger variation in the region denoting $68$ per cent of the sample. This most likely reflects the inclusion of a greater fraction of more massive objects in the CC profile, as the CC fraction increases with increasing mass for this criterion. Defining CCs by the cuspiness parameter results in the opposite trend, with NCC clusters having a marginally higher gas fractions at a fixed radius. The cuspiness parameter has a decreasing CC fraction with increasing mass, and fewer massive clusters are included in the CC profile.

\subsection{CC criteria distributions}
\begin{figure*}
 \includegraphics[height=20.3cm,width=1.04\textwidth]{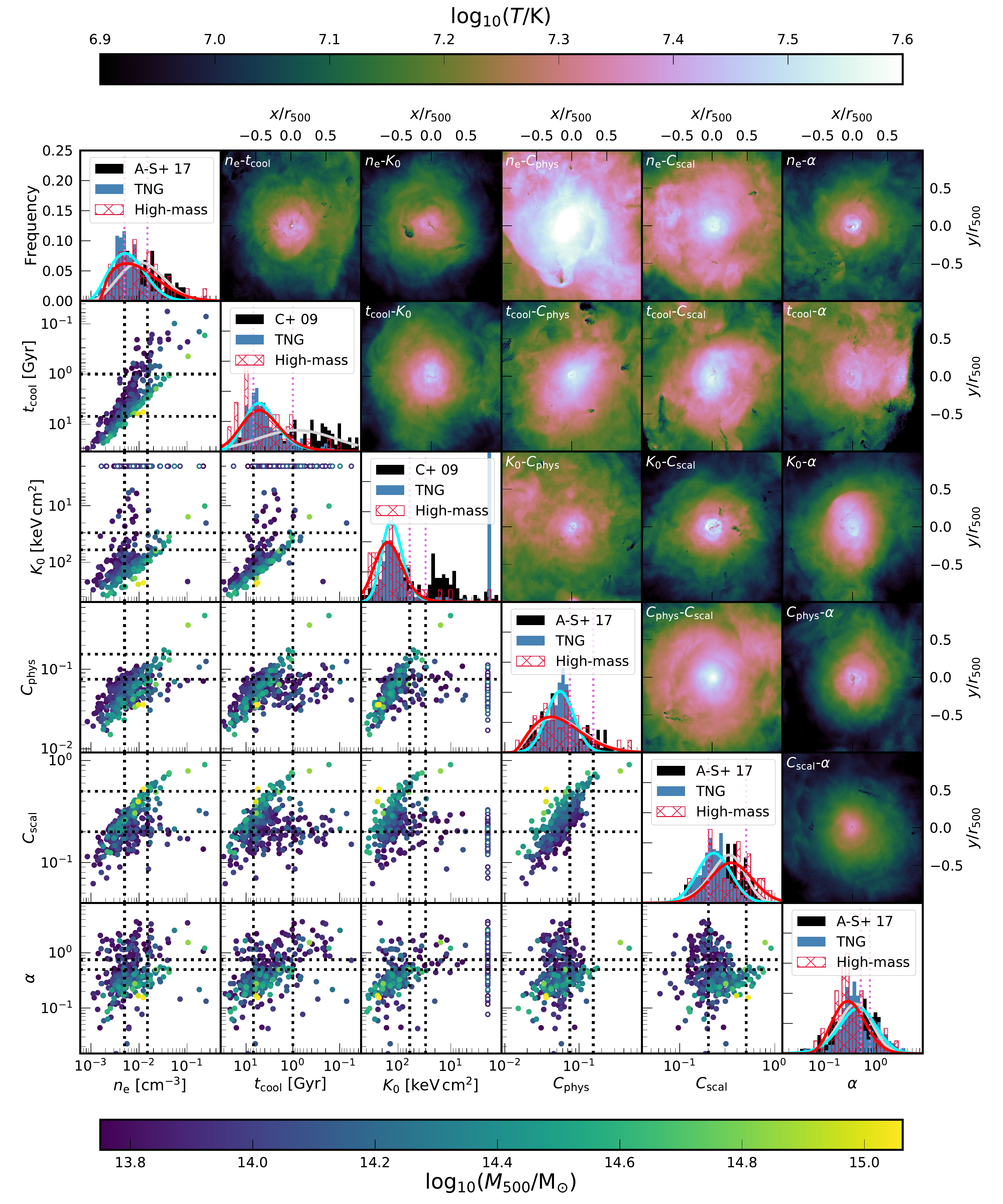}
 \caption{Comparison of the different CC criteria at $z=0$. In the diagonal panels we plot the complete sample (blue), the high-mass sample (red) and the observed sample (black) criteria distributions. We compare to the \textit{Planck} ESZ sample \citep{Andrade-Santos2017} [A-S+17] (for central electron number density, physical and scaled concentration parameter and cuspiness parameter) and the \textit{ACCEPT} sample \citep{Cavagnolo2009} [C+ 09] (for cooling time and central entropy excess). The solid blue, red and grey lines show the best-fit log-normal distributions for the complete sample, the high-mass sample and the observations, respectively. The lower-off diagonal panels show the simulated criteria correlations with color denoting cluster mass, where we find increased scatter for low-mass systems, and the dashed lines denoting CC and MCC thresholds. In the upper off diagonal panels we plot $2r_{500}\times2r_{500}$ mass-weighted temperature maps of randomly selected clusters that are defined as CC by both criteria.}
 \label{fig:cc_crit_comp}
\end{figure*}

We now compare the CC criteria distributions for the complete sample, the high-mass sample and the observational sample, which are plotted in the diagonal panels of Fig. \ref{fig:cc_crit_comp}. For the central electron number density, the concentration parameters and the cuspiness parameter we compare to \textit{Planck} ESZ sample \citep{Andrade-Santos2017} and for the central cooling time and central entropy excess we compare to the \textit{ACCEPT} cluster sample \citep{Cavagnolo2009}. Excluding the observed central entropy excess distribution, we find both the observed and simulated distributions are reasonably well described by log-normal distributions. The observed central entropy distribution is clearly bimodal. However, as noted, previous work by \citet{Panagoulia2014}, \citet{Hogan2017} and others have shown that the peak at $15\,\mathrm{keV}\,\mathrm{cm}^{2}$ appears to be generated by the limited resolution of the temperature profiles, which complicates a direct comparison.

A sizable fraction of the simulated clusters have a very low central entropy excess ($<2\,\mathrm{keV}\,\mathrm{cm}^{2}$), which may suggest that physical processes are missing from the IllutrisTNG model. All cosmological simulations currently lack the resolution and the physics to correctly capture the multi-phase nature of the ICM. Idealized simulations have shown a cluster's entropy floor is set by the ratio of the cooling time to the free-fall time \citep{McCourt2012,Sharma2012,Gaspari2012,Li2014}. Once $t_{\mathrm{cool}}/t_{\mathrm{ff}}\approx10$, cold gas begins to precipitate out of the hot gas, triggering AGN feedback events that maintain the central entropy. This result is supported by multiwavelength observations of filamentary molecular gas structures surrounding brightest cluster galaxies (BCGs) \citep{McDonald2010,McDonald2011,Werner2014,Tremblay2015,Suess2017}. Cosmological simulations may need to model AGN triggering by cold phase precipitation to reproduce the minimum entropy floor. However, observations \citep[e.g.][]{Panagoulia2014} have shown that the entropy profiles are almost pure power-laws at radii that are sufficiently well resolved, but the overall fraction of objects with these almost pure power-law profiles is uncertain as this study is X-ray selected and likely suffers from CC bias. If the clusters with negligible central entropy excess are removed, a comparison of the central entropy excess distributions for the complete and high-mass samples demonstrates that they are very similar, with mean values of $179\,\mathrm{keV}\,\mathrm{cm}^{2}$ and $176\,\mathrm{keV}\,\mathrm{cm}^{2}$ and standard deviations of $\sigma=0.42$ and $\sigma=0.52$ for the complete and high-mass samples, respectively.

The high-mass sample has significantly different central cooling time distribution compared to the observed distribution, but it is similar to the complete sample distribution. The complete and high-mass distributions have mean values of $6.29\,\mathrm{Gyr}$ and $5.42\,\mathrm{Gyr}$ and standard deviations of $\sigma=0.63$ and $\sigma=0.86$, respectively. In contrast, the observed distribution is very broad with a standard deviation of $\sigma=1.90$. Although selection effects, which we do not account for, will impact the observed distribution, the central cooling time is a balance between radiative losses, thermal conduction and heating by feedback processes and merger events. Therefore, it is unlikely the simulations will be able to reproduce the observed distribution without accurately modelling all of these processes.

\renewcommand\arraystretch{1.1}
\begin{table*}
 \caption{Correlation and scatter about the best-fit power law for the different CC criteria at $z=0$. The Spearman rank correlation coefficients, $r_{\mathrm{s}}$, are shown in the lower off-diagonal entries and the scatter, $\sigma_{\log_{10}}$, in the upper off-diagonal entries. Errors are computed by bootstrap resampling 10,000 times.}
 \begin{tabularx}{0.75\textwidth}{l R R R R R R}
 \hline
 Criterion & \multicolumn{1}{C}{$n_{\mathrm{e}}$} & \multicolumn{1}{C}{$t_{\mathrm{cool}}$} & \multicolumn{1}{C}{$K_{0}$} & \multicolumn{1}{C}{$C_{\mathrm{phys}}$} & \multicolumn{1}{C}{$C_{\mathrm{scal}}$} & \multicolumn{1}{C}{$\alpha$} \\
 \hline
 $n_{\mathrm{e}}$ & \multicolumn{1}{C}{\hspace{0.2cm}$-$} & $0.19\pm0.01$ & $0.35\pm0.03$ & $0.31\pm0.02$ & $0.36\pm0.03$ & $0.38\pm0.03$ \\
 $t_{\mathrm{cool}}$ & $0.88\pm0.04$ & \multicolumn{1}{C}{$-$} & $0.39\pm0.04$ & $0.51\pm0.03$ & $0.54\pm0.03$ & $0.49\pm0.04$ \\
 $K_{0}$ & $0.62\pm0.04$ & $0.86\pm0.02$ & \multicolumn{1}{C}{\hspace{0.2cm}$-$} & $0.73\pm0.02$ & $0.73\pm0.02$ & $0.63\pm0.03$ \\
 $C_{\mathrm{phys}}$ & $0.58\pm0.04$ & $0.49\pm0.05$ & $0.41\pm0.05$ & \multicolumn{1}{C}{\hspace{0.2cm}$-$} & $0.12\pm0.01$ & $0.18\pm0.01$ \\
 $C_{\mathrm{scal}}$ & $0.57\pm0.04$ & $0.27\pm0.05$ & $0.06\pm0.06$ & $0.74\pm0.03$ & \multicolumn{1}{C}{\hspace{0.2cm}$-$} & $0.18\pm0.01$ \\
 $\alpha$ & $0.31\pm0.06$ & $0.52\pm0.05$ & $0.62\pm0.04$ & $0.20\pm0.05$ & $-0.12\pm0.06$ & \multicolumn{1}{C}{\hspace{0.2cm}$-$} \\
 \hline
 \end{tabularx}
 \label{tab:correlations}
\end{table*}
\renewcommand\arraystretch{1.0}

The complete and high-mass samples have similar mean values for the central electron number density criterion, with values of $0.50\times10^{-2}\,\mathrm{cm}^{-3}$ and $0.58\times10^{-2}\,\mathrm{cm}^{-3}$, respectively. However, the high-mass sample has a larger standard deviation $\sigma=1.39$ compared to the complete sample $\sigma=0.97$ that is driven by a long tail towards higher central density values, producing the larger CC fraction. The observed sample has mean value of $1.07\times10^{-2}\,\mathrm{cm}^{-3}$, which is a factor $2$ larger than the high-mass sample and has a standard deviation of $\sigma=1.13$. Therefore, the observed CC fraction is likely $1\sigma$ larger than the observed fraction because the mean value of the distribution is higher, which is likely a consequence of simulations having lower than observed central gas fractions.

The concentration parameter distributions are the most discrepant distributions for the complete and high-mass samples. Within physical apertures they have mean values of $5.44\times10^{-2}$ and $3.92\times10^{-2}$ and standard deviations of $\sigma=0.43$ and $\sigma=0.99$, respectively. This compares to the observed distribution which has a mean value of $4.33\times10^{-2}$ and a standard deviation of $\sigma=0.99$. For the scaled concentration parameter the complete and high-mass samples have mean values of $0.22$ and $0.34$ and standard deviations of $\sigma=0.42$ and $\sigma=0.52$, respectively. The observed sample has a mean value of $0.51$ and standard deviation of $\sigma=0.26$. Therefore, although the high-mass sample has a lower mean value compared to the observed sample, the CC fraction is similar because the width of the high-mass sample distribution is larger.

The cuspiness parameter distributions of the complete sample and the high-mass sample are marginally different. The complete sample has a mean value of $0.46$ with a standard deviation of $\sigma=0.75$, while the high-mass sample has a lower mean value of $0.27$ and a similar standard deviation $\sigma=0.80$. Therefore, the complete sample has a higher CC fraction because the mean value of the distribution is larger. However, the simulated cuspiness parameter distribution is poorest fit by a log-normal distribution with a large number of clusters yielding values just below the MCC threshold, which leads to lower than observed CC fraction. The observational sample has a mean value of $0.41$ and a standard deviation of $\sigma=0.69$.

In the upper off-diagonal panels of Fig. \ref{fig:cc_crit_comp} we plot projected mass-weighted temperature maps of randomly selected simulated clusters that pass both CC criteria to demonstrate the range of systems that are classified as CCs. Each map is centred on the potential minimum of the cluster with a width of $2r_{500}\times2r_{500}$, a depth of $2r_{500}$ and is projected along the $z$-axis. All maps show a central core, but there is significant variation in the surrounding structures.

\subsection{Criteria correlations}
In the lower off-diagonal panels of Fig. \ref{fig:cc_crit_comp} we plot the correlation of the different CC criteria for the complete sample, with the point colour denoting $M_{500}$. We note that for easier comparison we have inverted the $t_{\mathrm{cool}}$ and $K_{0}$ axes, therefore CC clusters will always appear at the top or right side of a panel. In addition, we make those clusters with $K_{0}<2\,\mathrm{keV}\,\mathrm{cm}^{2}$ visible on the plots by setting $K_{0}=2\,\mathrm{keV}\,\mathrm{cm}^{2}$ for them. However, in quantifying the correlation coefficients of different CC criteria we use their fiducial values. We quantify the correlation between criteria using the Spearman rank correlation coefficient, $r_{\mathrm{s}}$. In addition, we fit a simple power-law of the form
\begin{equation}\label{eq:powerlaw}
 \log_{10}(Y) = A+B\log_{10}(X/X_{\mathrm{piv}})\,,
\end{equation}
where $A$ and $B$ set the normalisation and slope of the power-law, respectively, and $X_{\mathrm{piv}}$ is the pivot point, which is set to the median value of the criterion $X$. This enables us to compute the scatter about the best-fit via equation \ref{eq:scatter}. The correlation coefficients and scatter values for the complete sample are summarized in Table \ref{tab:correlations}, while the values for the high-mass sample are summarized in Appendix \ref{ap:cstab}.

First, we note that the correlation between criteria shows some mass dependence. Examining the $n_{\mathrm{e}}$-$\alpha$ correlation for the complete sample yields a correlation coefficient of $r_{\mathrm{s}}=0.31\pm0.06$, with a scatter of $\sigma_{\log_{10}} = 0.38\pm0.03$. Selecting clusters with $M_{500}>2\times10^{14.0}\,\mathrm{M}_{\astrosun}$ we find a significantly stronger correlation, with $r_{\mathrm{s}}=0.75\pm0.10$ and a scatter of $\sigma_{\log_{10}}=0.33\pm0.04$. This correlation coefficient is in better agreement with the observed value of $r_{\mathrm{s}}=0.88$ found for the \textit{Planck} ESZ sample \citep{Andrade-Santos2017}, which consists of clusters with $M_{500}>2\times10^{14}\,\mathrm{M}_{\astrosun}$ due to the selection function.

The central electron number density criterion, $n_{\mathrm{e}}$, is strongly correlated with $t_{\mathrm{cool}}$, $C_{\mathrm{phys}}$ and $C_{\mathrm{scal}}$, yielding correlation coefficients of $0.88\pm0.04$, $0.58\pm0.04$ and $0.57\pm0.04$, respectively. These correlations are not unexpected as the cooling time depends on the central number density, see equation (\ref{eq:cooling}), and clusters with higher central densities should have higher central X-ray luminosities, due to its $n_{\mathrm{e}}^{2}$ dependence. There is no obvious separation in the criteria space with cluster mass. The correlation of $t_{\mathrm{cool}}$ with the $C_{\mathrm{phys}}$ and $C_{\mathrm{scal}}$ criteria is significantly weakened by the increased scatter in these values for low-mass clusters, with correlation coefficients of $0.49\pm0.05$ and $0.27\pm0.05$ and scatter values of $0.51\pm0.03$ and $0.54\pm0.03$ for $t_{\mathrm{cool}}$-$C_{\mathrm{phys}}$ and $t_{\mathrm{cool}}$-$C_{\mathrm{scal}}$, respectively. Lower mass clusters systematically scatter to shorter central cooling times for a given concentration parameter.

The central entropy excess, $K_0$, is strongly correlated with $n_{\mathrm{e}}$, $t_{\mathrm{cool}}$ and $\alpha$, producing correlation coefficients of $0.62\pm0.04$, $0.86\pm0.02$ and $0.62\pm0.04$ respectively. There is a reasonable correlation with the concentration parameter measured within physical apertures with coefficient of $0.41\pm0.5$, but we find no correlation between $K_{0}$ and $C_{\mathrm{scal}}$ with a coefficient value of $-0.06\pm0.06$. Low-mass clusters have increased scatter to smaller values of $K_{0}$ for a given value of $C_{\mathrm{scal}}$ and many have a central entropy excess of zero, which results in no statistical correlation. We find a mild trend with mass for the $K_{0}$-$n_{\mathrm{e}}$ and $K_{0}$-$C_{\mathrm{scal}}$, with $n_{\mathrm{e}}$ and $C_{\mathrm{scal}}$ values increasing with mass for a given $K_{0}$ value.

We find a strong correlation between the physical and scaled concentration parameters, which is not that surprising. For the complete sample we find $r_{\mathrm{s}}=0.74\pm0.03$ with a small level of scatter of $\sigma_{\log_{10}}=0.12\pm0.01$ about the best-fit relation. We find a mass dependence for this correlation, with more massive clusters having a larger concentration parameter within physical apertures, $C_{\mathrm{phys}}$, for a given concentration parameter within scaled apertures, $C_{\mathrm{scal}}$. As noted above, this may be due to the distribution of gas with radius.

The cuspiness parameter, $\alpha$, is strongly correlated with $t_{\mathrm{cool}}$ $(r_{\mathrm{s}}=0.63\pm0.04)$ and $K_{0}$ $(0.60\pm0.04)$, and weakly correlated with $n_{\mathrm{e}}$ $(r_{\mathrm{s}}=0.33\pm0.06)$ and $C_{\mathrm{phys}}$ $(r_{\mathrm{s}}=0.23\pm0.06)$. However, it has some of the largest levels of scatter, with $t_{\mathrm{cool}}$ and $K_{0}$ producing values of $0.52\pm0.05$ and $0.62\pm0.04$ respectively.

In summary, we find that the correlation between different CC criteria is mass dependent, with an increasing correlation between criteria for more massive clusters. Low-mass clusters appear to scatter to lower central entropy excess values. Although correlated with other criteria the cuspiness parameter has some of the largest levels of scatter. The two quantities that are most correlated are the central electron number density and the central cooling time. 

\section{Evolution with redshift}
\label{sec:CCevo}
\subsection{CC fraction evolution}
\begin{figure*}
 \includegraphics[width=1.02\textwidth]{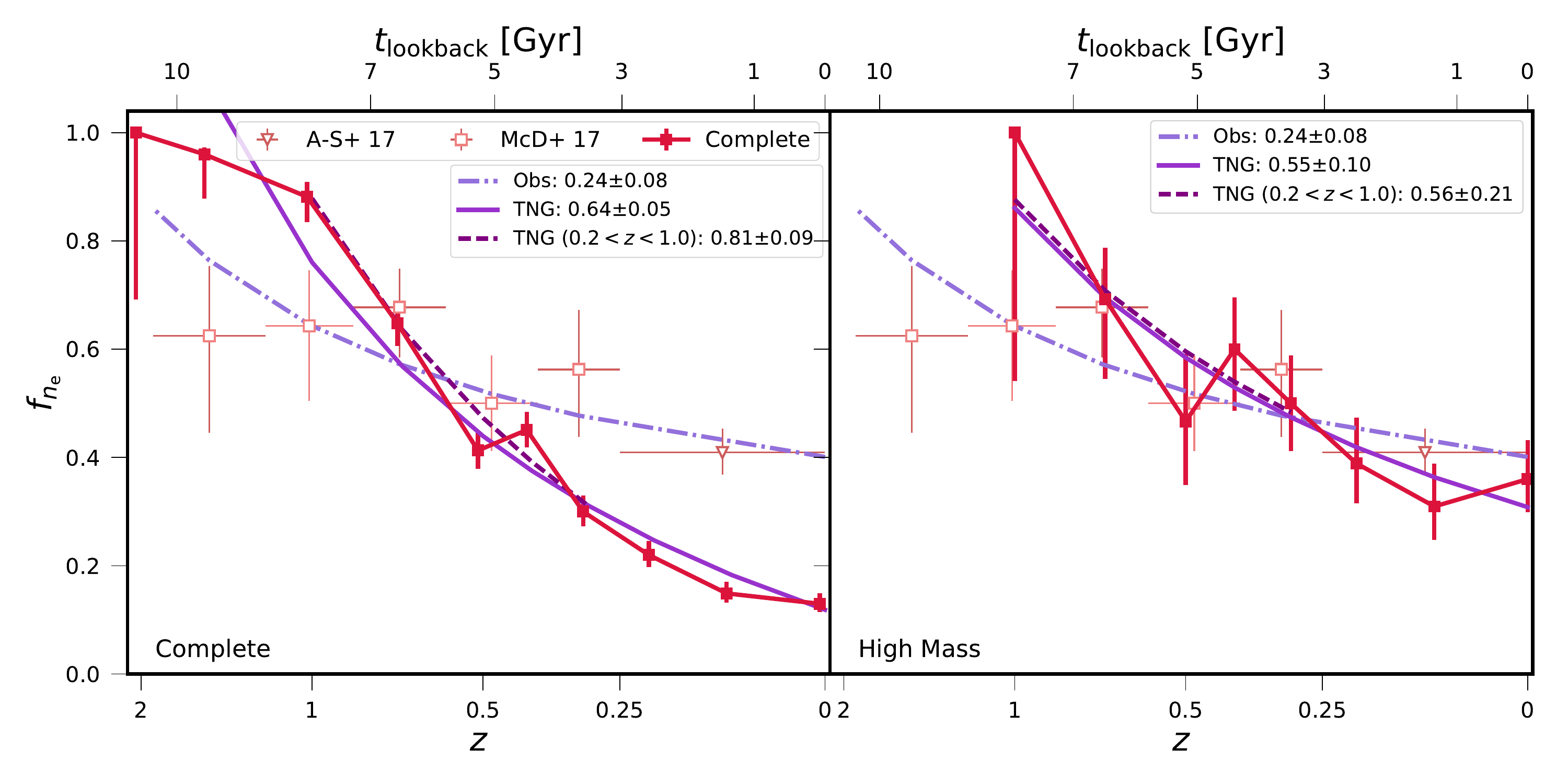}
 \caption{Evolution of the CC fraction, defined central electron number density, with redshift for the complete (left panel) and high-mass (right panel) samples (solid red line). We compare to the observed evolution for the same CC criterion from the combination of \citet{Andrade-Santos2017} (open circle) and \citet{McDonald2017} (open squares). The dash-dot, solid and dashed purple lines show the best-fit linear redshift evolution for the observed sample, the simulated sample and the simulated sample in the redshift range $0.2<z\leq1.0$, respectively, with the slopes given in the legend. The simulated evolution is consistent between the two different samples, but both samples are steeper than the observed evolution. Limiting the redshift range of the simulated samples results in a further steepening of the evolution.}
 \label{fig:CC_ne_zred}
\end{figure*}

\begin{figure*}
 \includegraphics[width=1.02\textwidth]{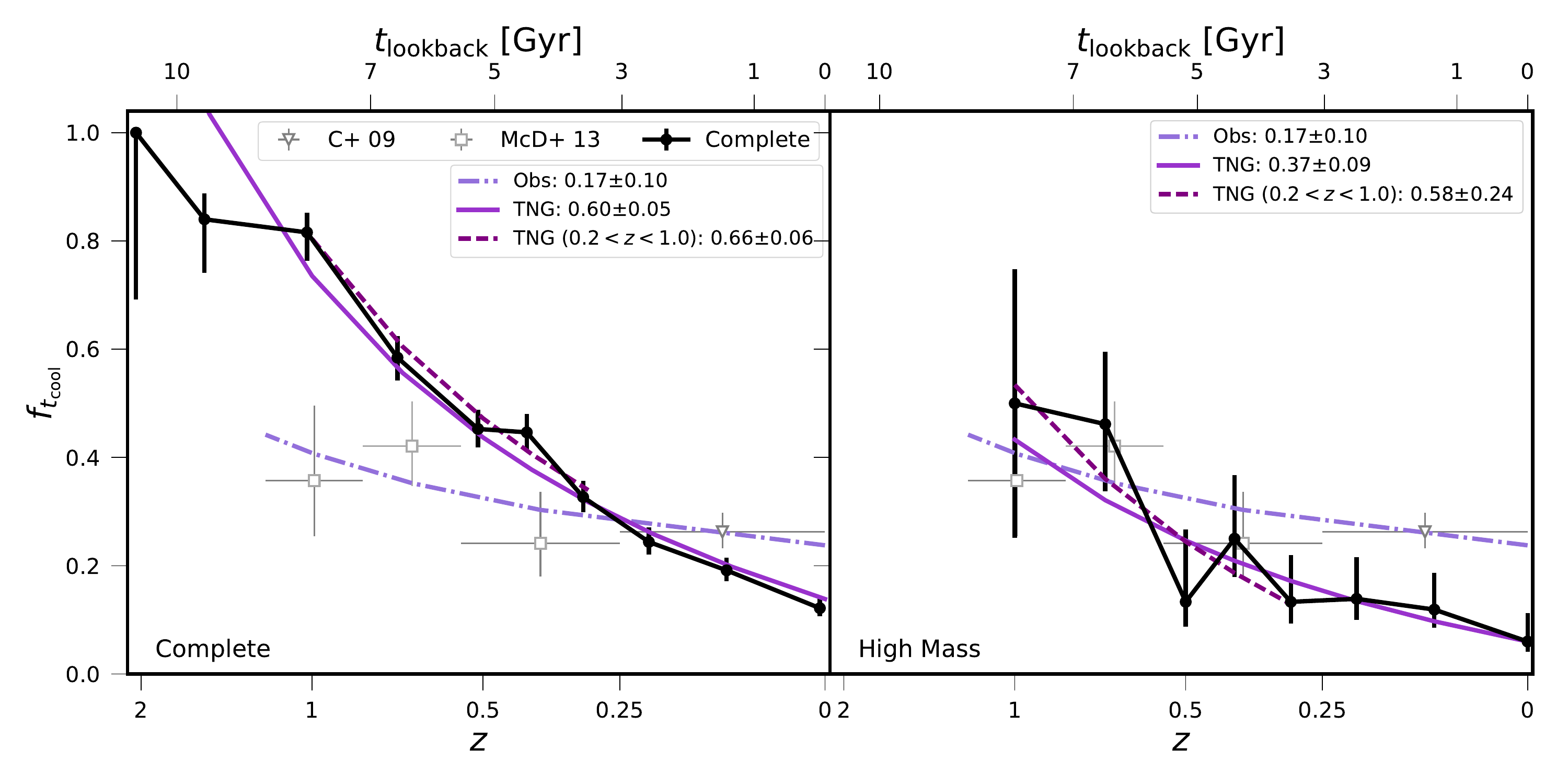}
 \caption{Evolution of the CC fraction with redshift for the complete (left panel) and high-mass (right panel) samples (solid black line), defined by the central cooling time criterion. We compare to the observed evolution from the combination of bias corrected \citet{Cavagnolo2009} (open circle) and \citet{McDonald2013} (open squares). The fit line styles are the same as in Fig. \ref{fig:CC_ne_zred}. The evolution with redshift for both simulated samples is steeper than the observed evolution.}
 \label{fig:CC_tcool_zred}
\end{figure*}

\begin{figure*}
 \includegraphics[width=1.02\textwidth]{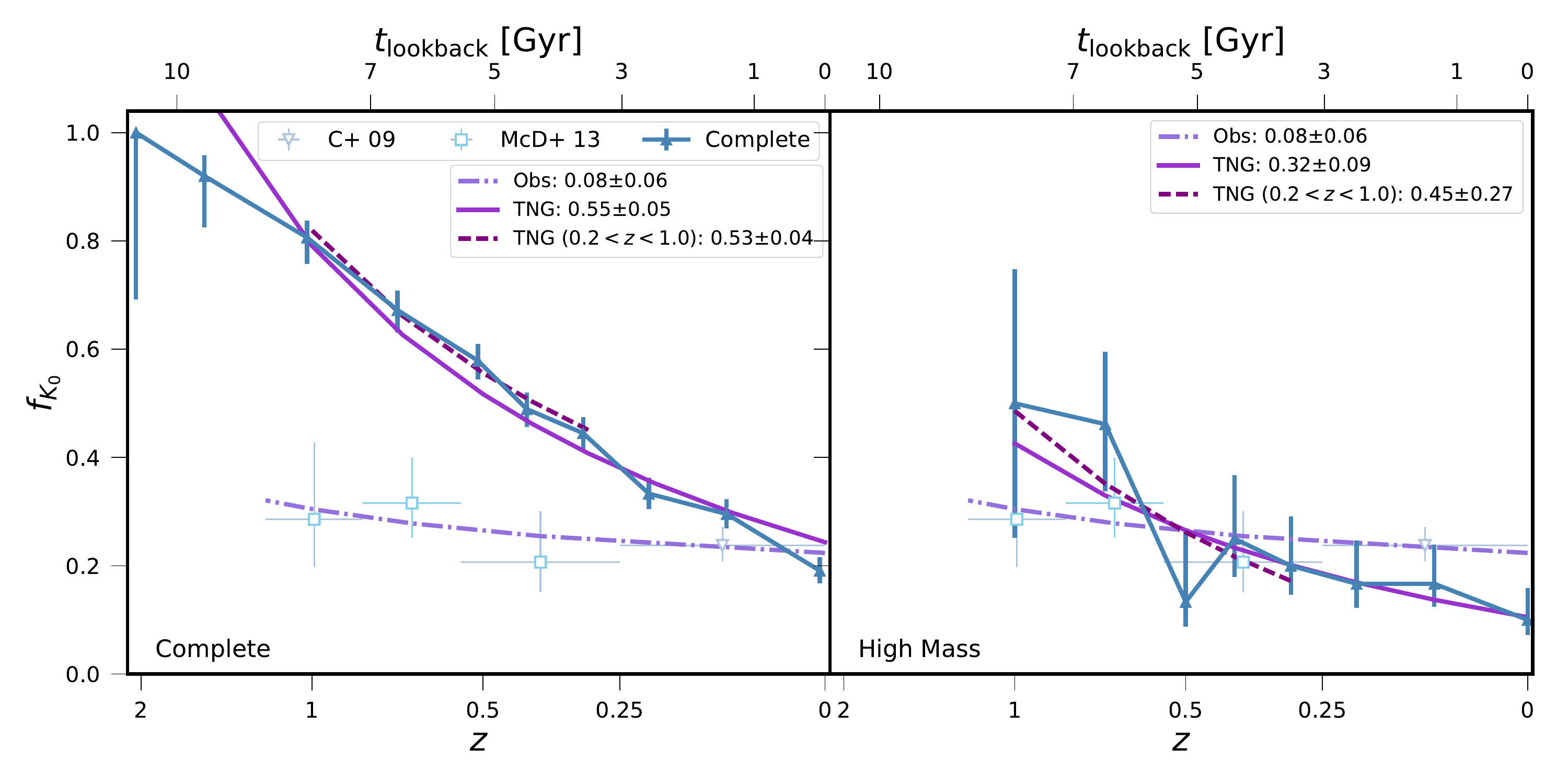}
 \caption{Evolution of the CC fraction with redshift for the complete (left panel) and high-mass (right panel) samples (solid blue line), defined by the central entropy excess criterion. We compare to the observed evolution from the combination of bias corrected \citet{Cavagnolo2009} (open circle) and \citet{McDonald2013} (open squares). The fit line styles are the same as in Fig. \ref{fig:CC_ne_zred}. The evolution with redshift for both simulated samples is steeper than the observed evolution.}
 \label{fig:CC_K0_zred}
\end{figure*}

\begin{figure*}
 \includegraphics[width=1.02\textwidth]{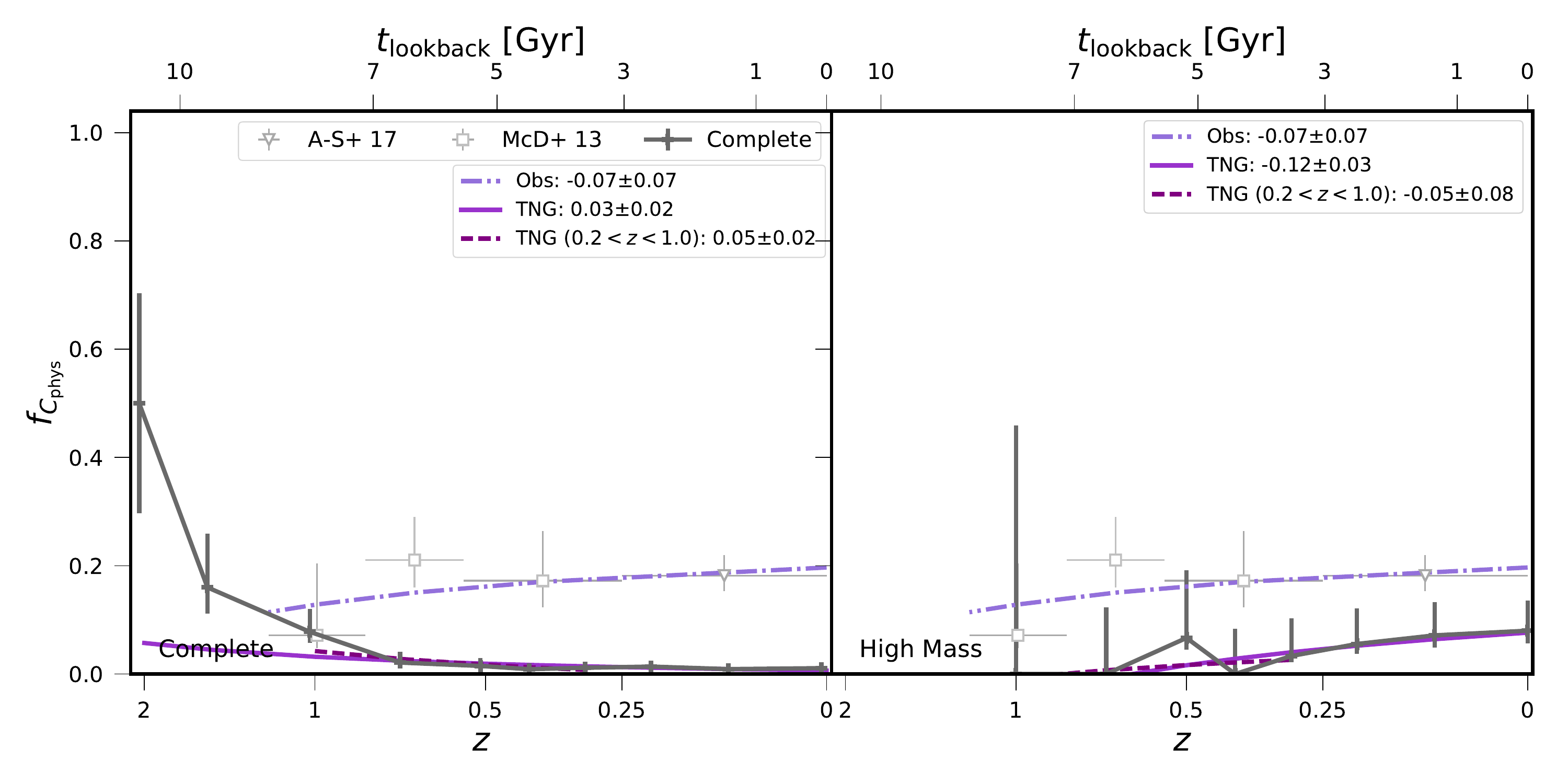}
 \caption{Evolution of the CC fraction with redshift for the complete (left panel) and high-mass (right panel) samples (solid grey line), defined by the concentration parameter within physical apertures. We compare to the observed evolution from the combination of \citet{Andrade-Santos2017} (open circle) and \citet{McDonald2013} (open squares). The fit line styles are the same as in Fig. \ref{fig:CC_ne_zred}. Selecting high-mass clusters results in a flatter and positive evolution with redshift that is more consistent with the observed evolution, however the normalization of the CC fraction is $0.15$ lower than observed.}
 \label{fig:CC_Cphys_zred}
\end{figure*}

\begin{figure*}
 \includegraphics[width=1.02\textwidth]{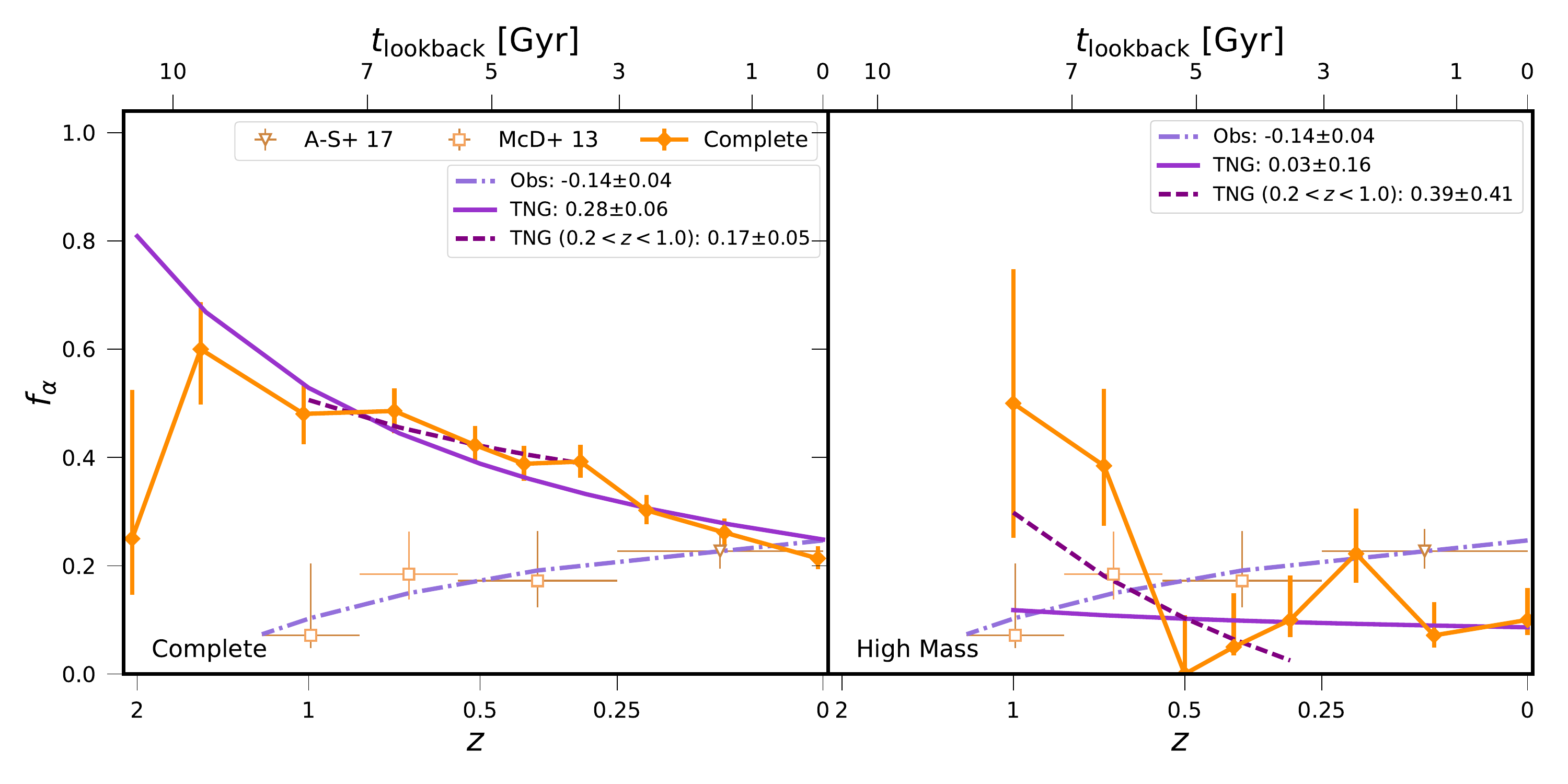}
 \caption{Evolution of the CC fraction with redshift for the complete (left panel) and high-mass (right panel) samples (solid yellow line), defined by the cuspiness parameter. We compare to the observed evolution from the combination of \citet{Andrade-Santos2017} (open circle) and \citet{McDonald2013} (open squares). The fit line styles are the same as in Fig. \ref{fig:CC_ne_zred}. The evolution with redshift for both simulated samples is consistent, however the evolution of the observed sample has a positive slope towards low-redshift and the simulated evolution is negative.}
 \label{fig:CC_alpha_zred}
\end{figure*}

Having examined the CC fraction at $z=0$ for the different criteria, we now compare the evolution of the CC fraction with redshift. The success of large SZ surveys with dedicated follow-up has resulted in a large number of clusters being detected out to $z\approx2$, due to the redshift independence of the SZ effect. This provides a large sample of clusters that is free from CC bias \citep{Lin2015} and enables a fair comparison of the redshift evolution of the simulated and observed CC fractions. We use the \textit{Planck} ESZ and \textit{ACCEPT} samples as a low-redshift ($z<0.25$) bin, with the latter's CC fractions corrected for the expected CC bias. The \textit{SPT-XVP} sample is then divided into $3$ bins over the redshift range $0.25<z<1.2$, with CC criteria values taken from \citet{McDonald2013}. For the central electron number density we additionally include the \textit{SPT-Hiz} sample to form a high-redshift bin $(1.2<z<1.9)$. The Planck ESZ and ACCEPT samples are both centred on the X-ray peak and for consistency we select the X-ray peak centred CC fractions for the SPT samples. \citet{McDonald2013} investigated the impact of centering choice and found no impact on the number of strong CCs found. In addition, for relaxed clusters the densest and strongest X-ray emitting gas will reside at the potential minimum and the centre of the simulated clusters is chosen to be the potential minimum. We note that the redshifts of SPT clusters \citep{Bleem2015} have been updated since \citet{McDonald2013} and this results in slightly different CC fractions compared to the original paper. The $1\sigma$ confidence intervals on both the observed and simulated samples are calculated via the beta distribution quantile technique \citep{Cameron2011}, due to low number statistics at high redshift. We note that \citet{McDonald2013} find similar low-redshift CC fractions to other low-redshift observations when analyzing the \citet{Vikhlinin2009a} cluster sample with the SPT pipeline, indicating that the systematic impact of using different observational samples at low-redshift is likely small.

\renewcommand\arraystretch{1.3}
\begin{table*}
 \caption{Table of the fraction of clusters defined as CC, MCC or NCC at $z=1$ for the criteria presented in Section \ref{sec:CCcrit} for the complete sample and the different mass bins. The $1\sigma$ uncertainties are computed by the beta distribution quantile technique.}
 \begin{tabularx}{\textwidth}{l C C C C C C C}
 \hline
 Sample & Fraction & \multicolumn{6}{c}{Criteria} \\
 $(z=1)$ & & $n_{\mathrm{e}}$ & $t_{\mathrm{cool}}$ & $K_{0}$ & $C_{\mathrm{phys}}$ & $C_{\mathrm{scal}}$ & $\alpha$ \\
 \hline
 Complete & CC & $0.87_{-0.05}^{+0.03}$ & $0.81_{-0.05}^{+0.04}$ & $0.83_{-0.05}^{+0.03}$ & $0.08_{-0.02}^{+0.04}$ & $0.00_{-0.00}^{+0.02}$ & $0.48_{-0.06}^{+0.06}$ \\
 $(M_{500}>10^{13.75}\,\mathrm{M}^{\astrosun})$ & MCC & $0.10_{-0.03}^{+0.05}$ & $0.18_{-0.04}^{+0.05}$ & $0.09_{-0.02}^{+0.04}$ & $0.45_{-0.05}^{+0.06}$ & $0.43_{-0.05}^{+0.06}$ & $0.16_{-0.03}^{+0.05}$ \\
 $N=77$ & NCC & $0.01_{-0.00}^{+0.03}$ & $0.00_{-0.00}^{+0.02}$ & $0.08_{-0.02}^{+0.04}$ & $0.47_{-0.06}^{+0.06}$ & $0.57_{-0.06}^{+0.05}$ & $0.36_{-0.05}^{+0.06}$ \\
 \hline
 Low-mass & CC & $0.87_{-0.07}^{+0.04}$ & $0.87_{-0.07}^{+0.04}$ & $0.93_{-0.06}^{+0.02}$ & $0.11_{-0.03}^{+0.06}$ & $0.00_{-0.00}^{+0.04}$ & $0.56_{-0.07}^{+0.07}$ \\
 $(M_{500}<9.0\times10^{13}\,\mathrm{M}_{\astrosun})$ & MCC & $0.09_{-0.03}^{+0.06}$ & $0.11_{-0.03}^{+0.06}$ & $0.02_{-0.01}^{+0.05}$ & $0.44_{-0.07}^{+0.07}$ & $0.33_{-0.06}^{+0.08}$ & $0.13_{-0.04}^{+0.07}$ \\
 $N=45$ & NCC & $0.02_{-0.01}^{+0.05}$ & $0.00_{-0.00}^{+0.04}$ & $0.04_{-0.01}^{+0.05}$ & $0.44_{-0.07}^{+0.07}$ & $0.67_{-0.08}^{+0.06}$ & $0.31_{-0.06}^{+0.08}$ \\
 \hline
 Intermediate-mass & CC & $0.87_{-0.09}^{+0.04}$ & $0.73_{-0.09}^{+0.06}$ & $0.70_{-0.09}^{+0.07}$ & $0.03_{-0.01}^{+0.07}$ & $0.00_{-0.00}^{+0.06}$ & $0.37_{-0.08}^{+0.09}$ \\
 $(9.0\times10^{13}\leq M_{500}<2.0\times10^{14}\,\mathrm{M}_{\astrosun})$ & MCC & $0.13_{-0.04}^{+0.09}$ & $0.27_{-0.06}^{+0.09}$ & $0.17_{-0.05}^{+0.09}$ & $0.50_{-0.09}^{+0.09}$ & $0.53_{-0.09}^{+0.09}$ & $0.20_{-0.05}^{+0.09}$ \\
 $N=30$ & NCC & $0.00_{-0.00}^{+0.06}$ & $0.00_{-0.00}^{+0.06}$ & $0.13_{-0.04}^{+0.09}$ & $0.47_{-0.09}^{+0.09}$ & $0.47_{-0.09}^{+0.09}$ & $0.43_{-0.08}^{+0.09}$ \\
 \hline
 High-mass & CC & $1.00_{-0.46}^{+0.00}$ & $0.50_{-0.25}^{+0.25}$ & $0.50_{-0.25}^{+0.25}$ & $0.00_{-0.00}^{+0.46}$ & $0.00_{-0.00}^{+0.46}$ & $0.50_{-0.25}^{+0.25}$ \\
 $(M_{500}\geq2.0\times10^{14}\,\mathrm{M}_{\astrosun})$ & MCC & $0.00_{-0.00}^{+0.46}$ & $0.50_{-0.25}^{+0.25}$ & $0.50_{-0.25}^{+0.25}$ & $0.00_{-0.00}^{+0.46}$ & $1.00_{-0.46}^{+0.00}$ & $0.00_{-0.00}^{+0.46}$ \\
 $N=2$ & NCC & $0.00_{-0.00}^{+0.46}$ & $0.00_{-0.00}^{+0.46}$ & $0.00_{-0.00}^{+0.46}$ & $1.00_{-0.46}^{+0.00}$ & $0.00_{-0.00}^{+0.46}$ & $0.50_{-0.25}^{+0.25}$ \\
 \hline
 \end{tabularx}
 \label{tab:CCfracs_z1}
\end{table*}
\renewcommand\arraystretch{1.0}

The complete sample has a lower median mass than the observational samples, due to observational selection functions and the limited volume of the simulation. Therefore, for each criterion we compare how the observed CC fraction evolves with redshift to both the complete and high-mass samples. At $z=1$, the mass bins contain $45$, $30$ and $2$ clusters and have median $M_{500}$ values of $6.8\times10^{13}\,\mathrm{M}_{\astrosun}$, $1.1\times10^{14}\,\mathrm{M}_{\astrosun}$ and $2.8\times10^{14}\,\mathrm{M}_{\astrosun}$, respectively. We summarize the fraction of clusters defined as CC, MCC and NCC at $z=1$ in Table \ref{tab:CCfracs_z1}. Additionally, we also examine the evolution only over the redshift range of the \textit{SPT-XVP} sample $(0.25<z<1.2)$ to ensure that differing low-redshift samples and low number statistics at $z>1.2$ do not impact the result. We do not examine the evolution of the concentration parameter with scaled apertures as there are no high-redshift ($z>0.25$) observational constraints. To enable a quantitative comparison we fit both the observed and simulated CC fraction as a function of redshift with a linear relation $(y=mx+c)$ using the Levenberg-Marquardt algorithm and accounting for the uncertainty in the CC fractions.

In Fig. \ref{fig:CC_ne_zred} we plot the CC fraction, defined by the central electron number density, for the complete sample (left panel) and the high-mass sample (right panel). As noted in the previous section the high-mass sample leads to higher normalization at low-redshift and the high-mass sample has a consistently greater CC fraction at fixed redshift. The CC fractions of the high-mass sample are in agreement with the observed, however the uncertainty is large due to the small sample size. To quantify the redshift evolution we a linear relation to the CC fraction as a function of redshift. With best-fit slopes of $0.64\pm0.05$ and $0.55\pm0.10$, respectively, the complete and high-mass samples have a consistent redshift evolution. This is significantly steeper than the observed evolution, which has a slope of $0.24\pm0.08$. If we only compare to the \textit{SPT-XVP} sample $(0.25<z<1.2)$ we find that the simulated slope steepens further for the complete sample. The observed CC fraction slowly decreases from $z=2$ to $z=0$. In contrast, the reduction in the CC fraction for the simulated samples begins later and results in a much more rapid decline. The high-mass sample begins to decline at lower redshift than the complete and then proceeds at the same rate, leading to a higher normalization.

We examine the evolution of the CC fraction defined by the central cooling time criterion with redshift in Fig. \ref{fig:CC_tcool_zred}. Though noisy and somewhat uncertain, there is tentative agreement between the high-mass CC fractions and the observations for $z>0.25$. At lower redshift the simulations have lower CC fractions. Fitting for the redshift evolution, we find that the complete sample produces a steeper slope of $0.60\pm0.05$ compared to the high-mass sample slope of $0.37\pm0.09$. The high-mass sample slope is $2\sigma$ steeper than the observed slope of $0.17\pm0.10$. If the considered redshift range is narrowed to $0.25<z<1.2$ the complete and high-mass samples yield consistent slopes that are steeper than observed.

Defining CCs via the central entropy excess yields very similar results to the central cooling time, as shown in Fig. \ref{fig:CC_K0_zred}. The high-mass sample produces CC fractions in reasonable agreement with the observed CC fractions, and a lower normalization relative to the complete sample. The discrepancy at low-redshift between the high-mass sample and the observed sample may be larger as the low-redshift observed CC fraction is likely a lower limit due to the limited resolution of the temperature profiles. Fitting the CC fractions we find the slope of the complete sample, $0.55\pm0.05$, is steeper than the slope of the high-mass sample, $0.32\pm0.09$. The high-mass sample slope is $>2\sigma$ steeper than the observed slope of $0.08\pm0.06$. When the redshift range is restricted the complete sample and the high-mass sample produce slopes that are consistent, but steeper than observed. We note that the limited observational resolution and a potentially higher low-redshift CC fraction would further flatten the observed redshift evolution of the CC fraction.

In Fig. \ref{fig:CC_Cphys_zred} we plot the observed and simulated CC fraction evolution with redshift for clusters defined as CCs by the concentration parameter within physical apertures. It is clear that the complete sample does not evolve linearly with redshift, producing a more rapid increase in CC fraction with increasing redshift. The linear fit to the complete sample CC fractions is therefore poor, but it yields a best-fit slope of $0.03\pm0.02$. The high-mass sample yields a slope of $-0.12\pm0.03$ that is in reasonable agreement with the observed slope of $-0.07\pm0.07$. However, the normalization of the simulated sample is consistently lower than the observed CC fraction at all redshifts by $\sim0.15$. Restricting the redshift range over which the linear relation is fit leads to negligible change in the slopes produced by the simulated samples.

Classifying clusters by the cuspiness parameter, as shown in Fig. \ref{fig:CC_alpha_zred}, we find that the complete sample yields a roughly linear redshift evolution, but the CC fraction decreases for the highest redshift bin. The complete sample yields a slope of $0.28\pm0.06$. The CC fraction of the high-mass sample is very noisy, with a lower CC fraction at low-redshift compared to the complete sample that generally increases with redshift. The best-fit slope of the high-mass sample is $0.03\pm0.16$, but is a comparatively poorer fit compared to the complete sample. The observed sample produces a slope of $-0.14\pm0.04$. If the redshift range of the comparison is restricted then both simulated samples produce a positive slope with increasing redshift, but the trend for the high-mass sample is very uncertain and still consistent with no redshift evolution.

If the entire cluster volume evolves in a self-similar manner the properties of the core should change with redshift and this would result in the evolution of the CC fraction with redshift. Clusters are defined as overdensities relative to the critical density of the Universe, which increases with increasing redshift. This results in clusters of a fixed mass increasing in density with increasing redshift. Measuring the electron number density at fixed fraction of a radius relative to this overdensity will result in it being measured closer to the cluster centre with increasing redshift and it will increase. This results in a greater fraction of clusters being defined as CC for a fixed physical threshold. From equation (\ref{eq:cooling}), it is clear that an increasing number density will result in a decreasing cooling time and for a fixed threshold the fraction of clusters defined as CC via their central cooling time should increase with redshift. The concentration parameter within physical apertures should yield a greater CC fraction with redshift because the inner aperture will include an increasing fraction of the X-ray emission and the outer aperture will remain roughly unchanged due to the $n_{\mathrm{e}}^{2}$ dependence of the emission. The properties of cluster cores are observed to remain relatively unchanged with redshift \citep{McDonald2017}, while the rest of the cluster volume follows the expected self-similar expectation. Therefore, differences in the observed and simulated slope, i.e. the redshift evolution, may by driven by differences in how the thermodynamic properties of the core change with redshift.

\begin{figure*}
 \includegraphics[width=1.02\textwidth]{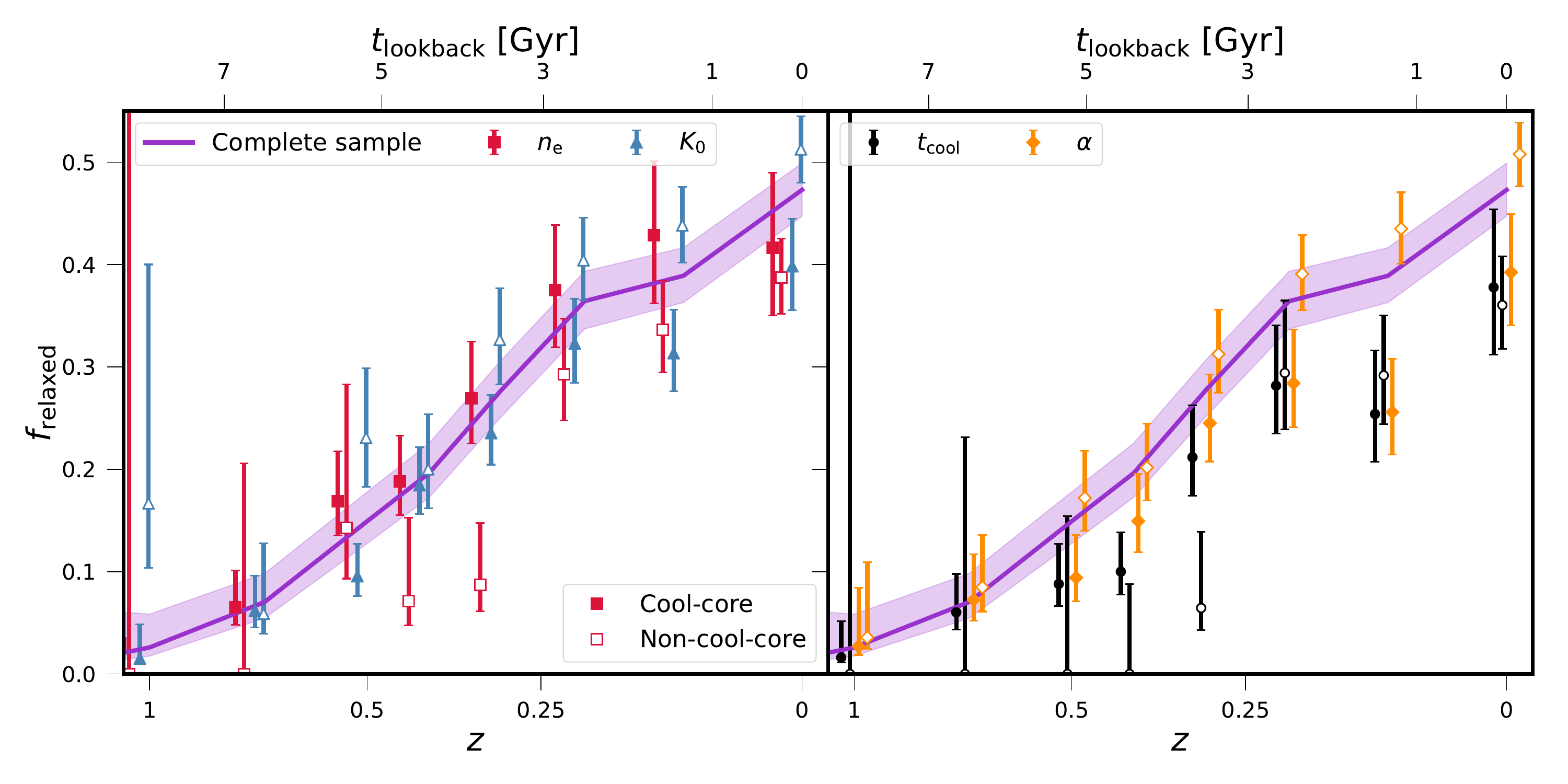}
 \caption{Fraction of clusters defined as relaxed as a function of redshift. We plot the relaxed fraction for the complete sample (purple line),  those complete sample clusters defined as CCs (filled symbols) and those complete sample clusters defined as NCC (open symbols). CCs are defined via the central electron number density (red square), cooling time (black circle), central entropy excess (blue triangle) and the cuspiness parameter (yellow diamond) criteria. The shaded region and error bars denote $1\sigma$ confidence intervals, calculated via the beta distribution quantile technique. The CC and NCC points are divided between the two panels and marginally offset for clarity. We find no conclusive evidence that a greater fraction of CCs, defined by any criteria, are defined as relaxed compared to the NCC or complete samples.}
 \label{fig:Relaxation}
\end{figure*}

\subsection{Relaxed fraction}
CC clusters are often associated with more spherical and regular X-ray morphologies. Previous numerical work has shown that close to head-on major mergers can disrupt a CC \citep{Rasia2015,Hahn2017}, which suggests that a greater fraction of NCC clusters should be disturbed at low-redshift. Additionally, radio haloes that are associated with recent merger activity have, to date, only been observed in NCC clusters, which suggests that a greater fraction of NCC clusters should have undergone recent mergers \citep{Cassano2010}. We now examine the fraction of clusters that are defined as relaxed and the fraction of CC and NCC clusters that are defined as relaxed as a function of redshift. Theoretically, there are many ways of defining a relaxed cluster \citep{Neto2007,Duffy2008,Klypin2011,DuttonMaccio2014,Klypin2016}. We follow \citet{Barnes2017b} and define a cluster as relaxed if 
\begin{equation}
 E_{\mathrm{kin},500}/E_{\mathrm{therm},500} < 0.1\,,
\end{equation}
where $E_{\mathrm{kin},500}$ is the sum of the kinetic energy of the gas cells, with the bulk motion of the cluster removed, inside $r_{500}$. This should account for any motions generated by substructures or the centre of mass being offset from the potential minimum. $E_{\mathrm{therm},500}$ is the sum of the thermal energy of the gas cells within $r_{500}$. As clusters relax they will thermalize, converting kinetic energy to thermal energy via weak shocks \citep[e.g.][]{Kunz2011} and potentially turbulent cascades \citep{Zhuravleva2014}. We demonstrate in Appendix \ref{app:rlx} that using other metrics to define a relaxed cluster, such as substructure fraction or centre of mass offset, yield similar results. We note that the selected threshold for the ratio of kinetic to thermal energy is designed to yield a relaxed sample, as opposed to other criteria that select thresholds to remove the most disturbed objects.

In Fig. \ref{fig:Relaxation} we plot the relaxed fraction as a function of redshift in both panels, where we have split the criteria for clarity. The fraction of the complete sample defined as relaxed decreases with increasing redshift, from $45\pm3$ per cent at $z=0$ to $3\pm1$ per cent at $z=1$. The increased kinetic energy of the cluster gas with increasing redshift has been shown in previous numerical work \citep{Stanek2010,Barnes2017a,LeBrun2017}, which is consistent with the picture that the merger rate at higher redshift is larger \citep{McBride2009,Fakhouri2010,Giocoli2012} and that clusters have had less time to thermalize at high-redshift. This picture is consistent with observational results if the crossing time of clusters decreases with increasing redshift \citep{Mantz2015,Nurgaliev2017,McDonald2017}, which it does if we assume self-similar evolution \citep{Carlberg1997}.

In addition, we plot the fraction of complete sample that is defined as a CCs or NCCs that are classified as relaxed, via the central electron number density and the central entropy excess in the left panel, and the central cooling time and cuspiness parameter in the right panel. We do not show the concentration parameter because it is noisy as it only defines a small number of clusters as CC. We are limited to $z\leq1$ due to small number statistics at high redshift. Defining CCs and NCCs by the central electron number density, we find that the fraction of relaxed clusters in both samples is broadly consistent with the complete sample, decreasing from $42\pm7$ per cent and $39\pm4$ per cent at $z=0$ to $13\pm5$ per cent and $14^{+14}_{-5}$ per cent at $z=0.5$ for the CC and NCC samples respectively. For $z>0.5$ no clusters are defined as NCC. We find that the fraction of CCs defined as relaxed is consistent within $2\sigma$ of the fraction of NCCs defined as relaxed at all redshifts. Defining clusters as CC or NCC by their central entropy excess, there is some evidence that the fraction of NCC clusters defined as relaxed is greater than the fraction of CC clusters defined as relaxed, but the fractions are consistent within $2\sigma$. We find that $37\pm6$ per cent and $49\pm3$ per cent at $z=0$ and $10\pm3$ and $23\pm5$ per cent at $z=0.5$ of clusters are defined as relaxed for the CC and NCC samples, but we note that many CCs have a central entropy excess of zero. For $z>0.5$ no clusters are defined as NCC.

\begin{figure*}
 \includegraphics[width=1.02\textwidth]{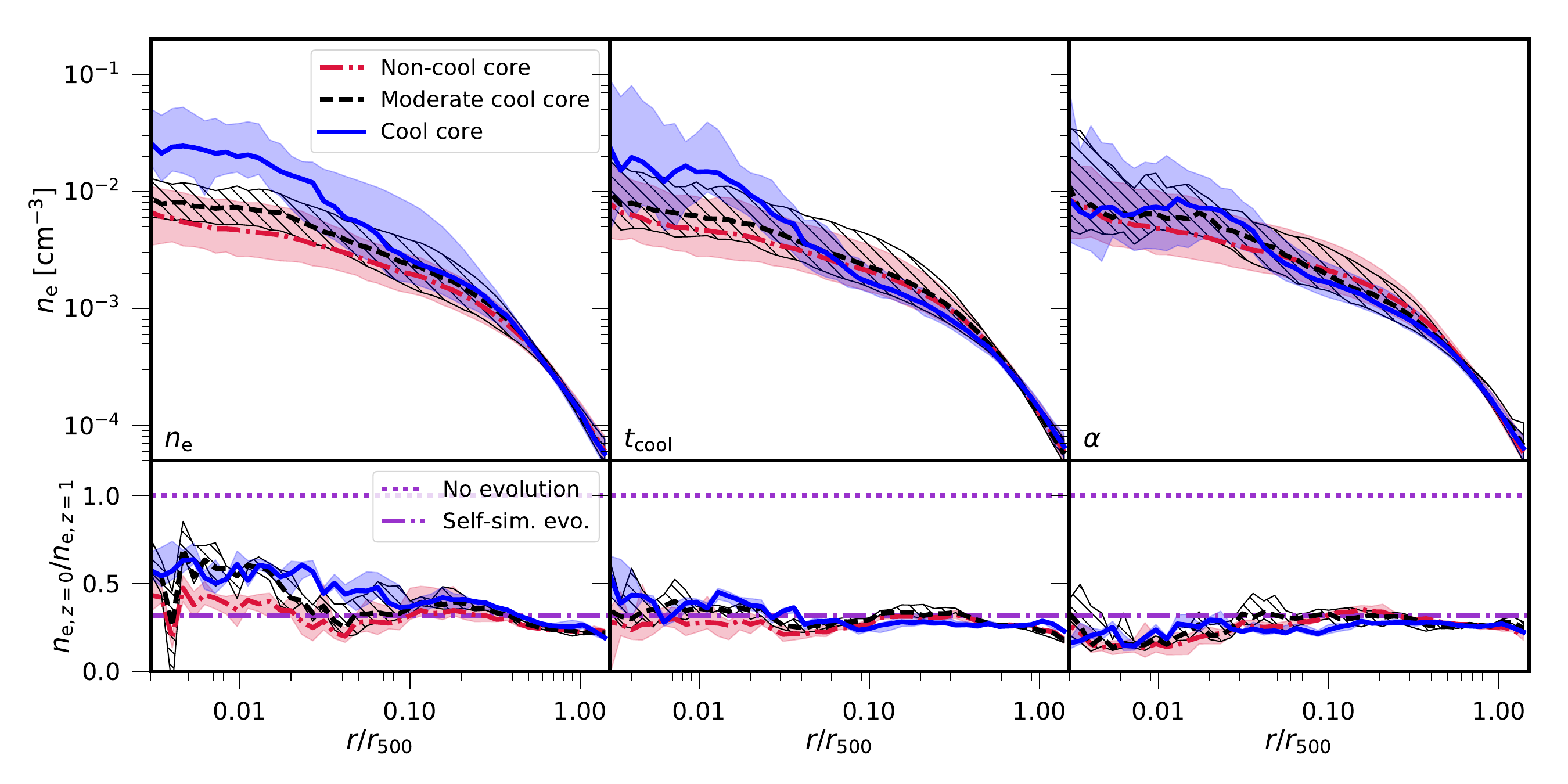}
 \caption{Median electron number density profiles at $z=0$ (top row) for clusters that are defined as CC (blue solid line), MCC (dashed black line) and NCC (dash-dot red line) according to the central electron number density (left panel), cooling time (centre panel) and cuspiness parameter (right panel). Filled and hatched regions enclose $68$ per cent of the sample. In the bottom panels we plot the ratios of the median profiles at $z=0$ and $z=1$, with the purple dotted and dash-dot lines denoting self-similar and no evolution, respectively. All criteria show at least some evolution, in contrast to the observations that show the core region does not evolve between $z=1.5$ and $z=0$.}
 \label{fig:rho_profiles}
\end{figure*}

Using central cooling time as the defining criteria, we find that the fraction of the CC sample defined as relaxed is consistent with the relaxed fractions for the NCC and the complete samples. At $z=0$, $38\pm7$ per cent of CC clusters are defined as relaxed, compared to $36\pm5$ per cent for the NCC sample. In contrast, defining CCs via the cuspiness parameter we find that at low-redshift $(z<0.25)$ the fraction of CC clusters defined as relaxed is lower than for the NCC sample. We find that $41\pm5$ per cent and $50\pm3$ per cent of CCs and NCCs, respectively, are defined as relaxed at $z=0$. At higher redshifts the fraction of CCs and NCCs defined as relaxed are consistent with each other and the relaxed fraction of the complete sample. Overall, we find little evidence that the simulated CCs samples have a higher relaxed fraction compared to either the NCC sample or the complete sample for any of the CC criteria that we have examined. We stress this result holds for other theoretical methods of defining a relaxed cluster, such as substructure fraction and centre of mass offset. For the central entropy excess and the cuspiness parameter criteria the fraction of NCC clusters defined as relaxed is larger than fraction of CC clusters defined as relaxed. However, the difference is marginal and we would require a significantly larger sample to investigate this further.

In summary, we find that the simulated CC fraction evolves significantly more with redshift than the observed fraction. The process that converts CCs to NCCs appears to begin later in the simulations but acts much more rapidly. We find no evidence that the relaxed fraction of CCs is greater than NCCs or the complete cluster sample, suggesting that mergers are not solely responsible for disrupting CCs. We stress that, as shown in Appendix \ref{app:rlx}, these results hold for other theoretical methods of defining a relaxed cluster.

\section{Core evolution}
\label{sec:profs}
We now examine how the properties of the average simulated cluster core evolves for CC, MCC and NCC clusters. \citet{McDonald2017} demonstrated that the cores of CC clusters present in the SPT samples have evolved little since $z\geq1.5$, with the rest of the cluster volume evolving in a manner consistent with the self-similar expectation. All profiles are computed in the range $10^{-3}-1.5\,r_{500}$ using $50$ radial bins. Only non-star forming gas that is cooling, i.e. not being heated via supernovae or AGN feedback, with a temperature $T>1.0\times10^{6}\,\mathrm{K}$ is included in the profiles. We use the central electron number density, central cooling time and cuspiness parameter as CC criteria in this section. Both concentration parameters yield very few CCs at $z=0$, which makes the median profiles too noisy, while the central entropy excess yields a large number of clusters with a negligible central entropy which may be the result of missing physical processes. Where appropriate we normalize by the profiles at $z=0$ by the expected virial quantity of the cluster. These quantities are defined as
\begin{equation}
 k_{\mathrm{B}}T_{500}=\frac{GM_{500}\mu m_{\mathrm{p}}}{2r_{500}}\:,
\end{equation}
\begin{equation}
 P_{500}=500f_{\mathrm{b}}k_{\mathrm{B}}T_{500}\frac{\rho_{\mathrm{crit}}}{\mu m_{\mathrm{p}}}\:,
\end{equation}
\begin{equation}
 K_{500}=\frac{k_{\mathrm{B}}T_{500}}{(500f_{\mathrm{b}}\rho_{\mathrm{crit}}/\mu_{\mathrm{e}}m_{\mathrm{p}})^{2/3}}\:,
\end{equation}
where $k_{\mathrm{B}}$ is the Boltzmann constant, $G$ is the gravitational constant, $\mu=0.59$ is the mean molecular weight, $m_{\mathrm{p}}$ is the proton mass, $\rho_{\mathrm{crit}}\equiv E^{2}(z)(3H_{0}^{2}/8\mathrm{\pi}G)$, $H_{0}$ is the Hubble constant and $E(z)\equiv(\Omega_{\mathrm{M}}(1+z)^{3}+\Omega_{\Lambda})^{1/2}$.

\subsection{Electron number density profiles}
In Fig. \ref{fig:rho_profiles} we plot the median electron number density profiles at $z=0$ in the top row of panels.  We divide the clusters into CC, MCC and NCC samples via the central electron number density, central cooling time and cuspiness parameter in the left, centre and right panels respectively. The shaded/hatched region denotes the region the encompasses $68$ per cent of the sample. The median CC, MCC and NCC profiles defined by the central electron number density criterion are in good agreement for $r>0.1r_{500}$, but the CC profile diverges inside of this radius. The median CC central density is a factor $\approx2.5$ greater than the median MCC profile, which in turn is $30$ per cent higher than the median NCC profile. Dividing clusters by their central cooling time the profiles begin to diverge at $r=0.1r_{500}$, reach a central density that is a factor $2$ lower than central electron number density criterion, and have a significantly larger spread. When the sample is classified by the cuspiness parameter we find that the CC, MCC and NCC median profiles are all consistent with each other within the $1\sigma$ uncertainties at all radii.

In the bottom row of panels we plot the ratio of the median CC, MCC and NCC profiles at $z=0$ over the median CC, MCC and NCC profiles at $z=1$. Note the samples are redefined at $z=1$ and clusters can change from a CC to NCC, vice versa, or disappear from the sample between the two redshifts, for example by dropping below the mass threshold of $M_{500}>10^{13.75}\,\mathrm{M}_{\astrosun}$ at high-redshift. If the profile has not evolved we would expect the ratio to be unity, i.e. $n_{\mathrm{e}}^{z=0}(r/r_{500})=n_{\mathrm{e}}^{z=1}(r/r_{500})$. The self-similar model \citep{Kaiser1986} is scale free and so all overdensities will evolve in the same manner and the ratio would produce a value of $1/E^{2}(z=1)=0.316$. The uncertainty in the ratio is calculated by bootstrap resampling the $z=0$ and $z=1$ samples $10,000$ times. All profiles evolve in reasonable agreement with the self-similar expectation for $r>0.1r_{500}$. Inside this radius the CC ratio diverges from the NCC ratio when separating clusters by their central electron number density. The CC ratio begins to rise at $r=0.1r_{500}$ and reaches a peak value of $0.60$, while the MCC ratio begin rising at $r=0.03r_{500}$ and reaches a peak value of $0.60$. The NCC ratio reaches a peak value of $0.42$. In contrast, using the central cooling time or the cuspiness parameter as the CC criterion results in all ratios being consistent with expected self-similar evolution at all radii, except at $0.01r/r_{500}$ where the ratio is less than the self-similar expectation for the cuspiness parameter. We conclude that the extent to which the density of the typical cluster core has evolved is dependent on the CC criteria used. For the central electron number density the profiles have evolved more than observed since $z=1$ \citep{McDonald2017}, but less than the self-similar expectation.

\subsection{Temperature profiles}
\begin{figure*}
 \includegraphics[width=1.02\textwidth]{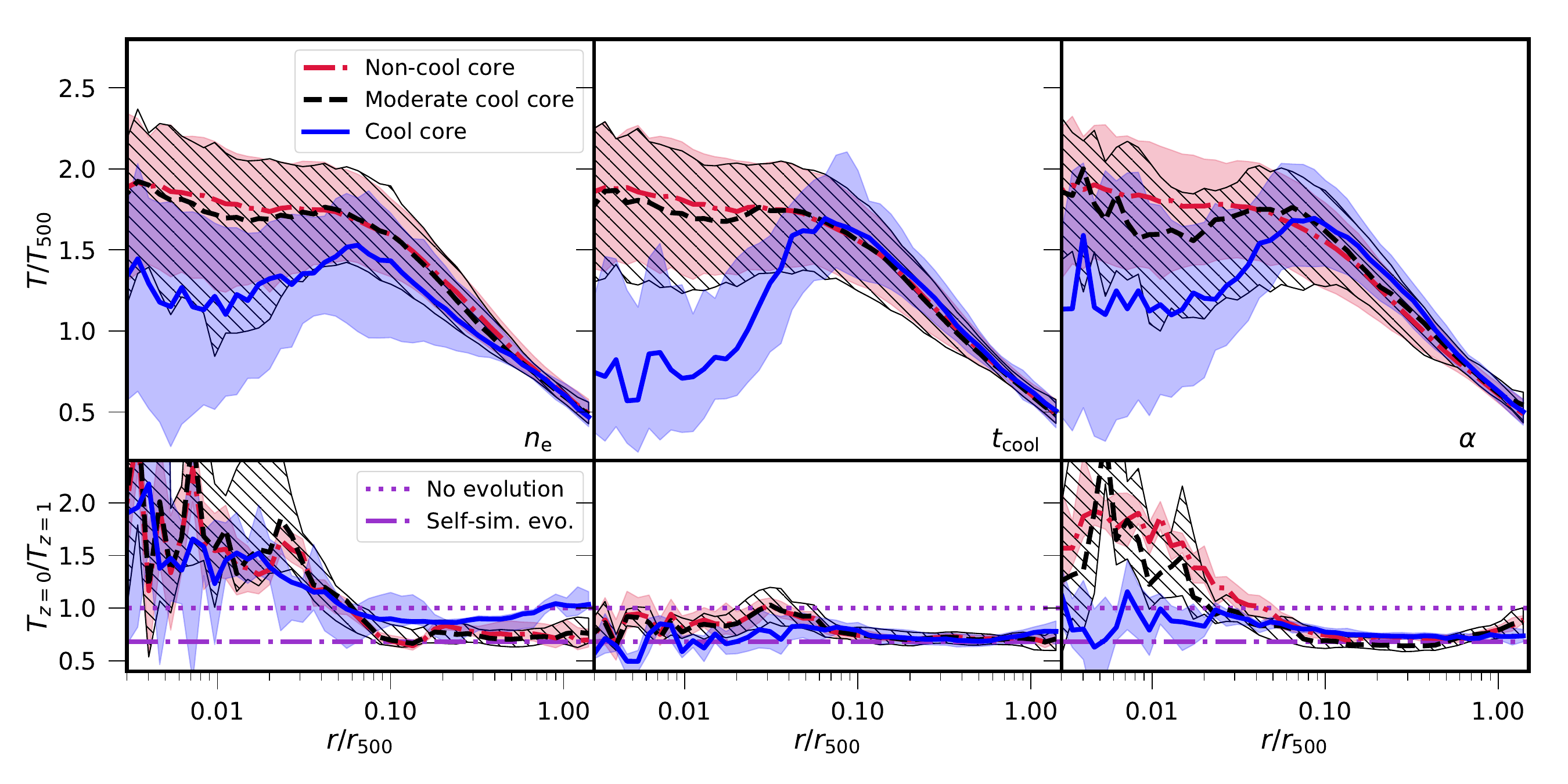}
 \caption{Median gas temperature profiles at $z=0$ (top row) and the ratio of the median temperature profiles at $z=0$ and $z=1$ (bottom panel). The line styles are the same as in Fig. \ref{fig:rho_profiles}. For the central electron number density and cuspiness parameter criteria the temperature in the core of all clusters deviates from the self-similar expectation.}
 \label{fig:T_profiles}
\end{figure*}

\begin{figure*}
 \includegraphics[width=1.02\textwidth]{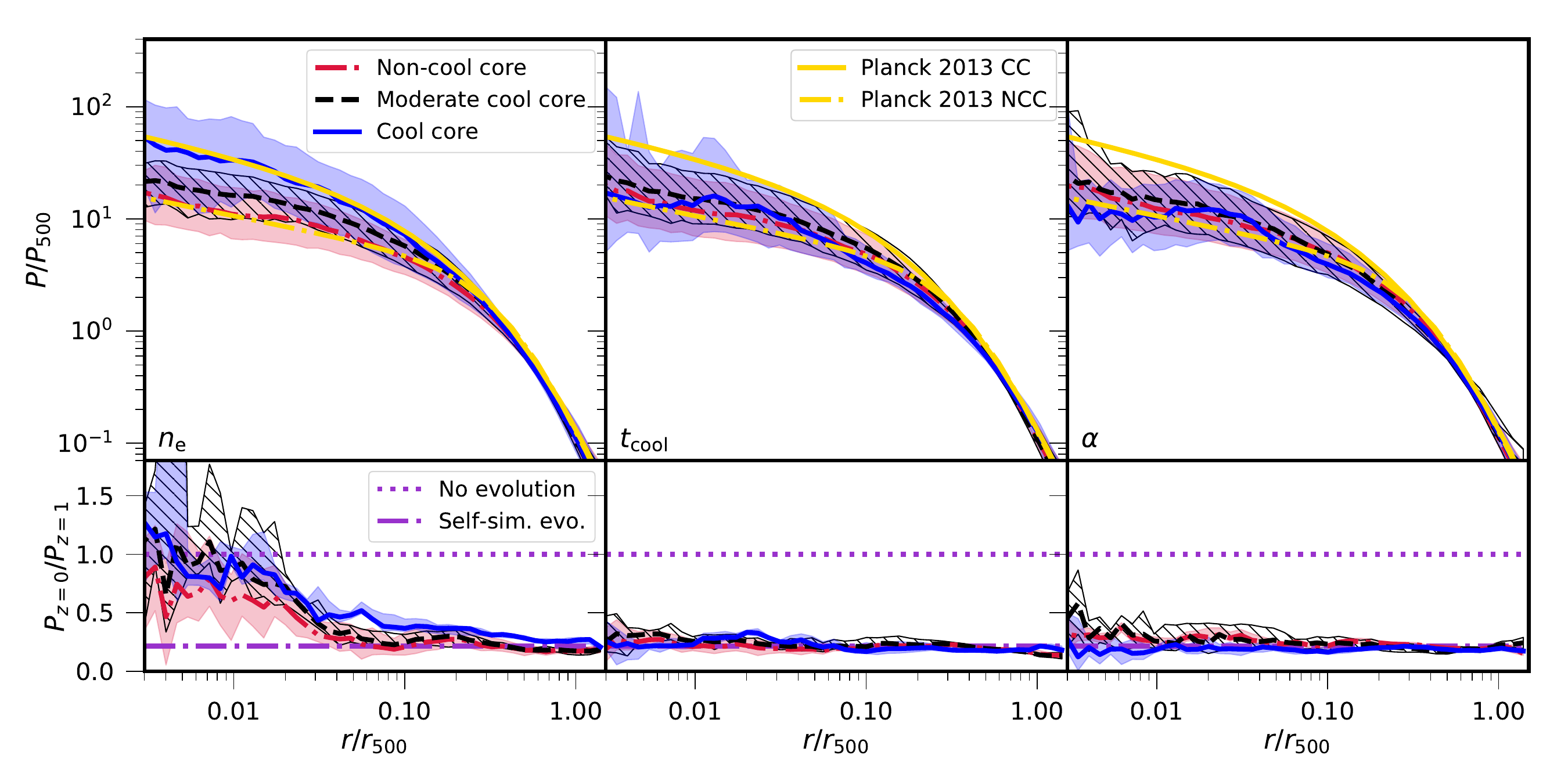}
 \caption{Median gas pressure profiles at $z=0$ (top row) and the ratio of median pressure profiles at $z=0$ and $z=1$ (bottom row). The line styles are the same as in Fig. \ref{fig:rho_profiles}. We plot the observed universal pressure profile (yellow) for the CC (solid) and NCC (dot-dash) clusters using the median mass for each simulated sample \citep{Planck_IR_V}. Using the same criterion the observed and simulated profiles show good agreement, but the differing mass dependence of other criteria results in different CC profiles.}
 \label{fig:P_profiles}
\end{figure*}

We plot the median CC, MCC, and NCC temperature profiles at $z=0$ in the top row of Fig. \ref{fig:T_profiles}. For $r>0.1r_{500}$ the median CC, MCC and NCC profiles are consistent with each other, within the scatter, for all CC criteria. Inside this radius, the central cooling time criterion produces a CC profile that reduces to a third of the peak value, while the MCC and NCC profiles are flat to the centre. Classifying clusters by the cuspiness parameter produces a central CC profile that reduces by a third, a MCC profile that shows a modest reduction and a flat NCC profile. The central electron number density criterion produces a CC profile that shows a mild reduction in the central temperature, with the MCC and NCC profiles flattening in the core. The CC profile shows significantly more scatter compared to other criteria. The increased scatter in the CC profile is due to the fact that the criteria has a consistent number of clusters in each mass bin, which results in a greater spread in the normalization of the temperature profile as more massive clusters are hotter due to their deeper potential wells. The other criteria preferentially select haloes from one mass bin, which makes the normalization of their temperature profiles more consistent. All profiles begin to rise again at $\sim0.005r_{500}$, which may be a sign of ongoing AGN activity. 

The bottom panels show the ratio of median temperature profiles at $z=0$ and $z=1$. If the profile has evolved self-similarly this ratio would be $1/E^{2/3}(z=1)=0.681$. For $r\geq0.1r_{500}$ all profiles are consistent with the expected self-similar evolution. The exception is the central electron number density CC profile which has cooled less than expected, but we note this profile has a very large scatter at $z=0$. Inside this radius the ratios defined by the central electron number density and cuspiness parameter criteria rise to values consistent with or greater than no evolution, suggesting the cores have gotten hotter since $z=1$. In contrast, if the clusters are classified by the central cooling time or cuspiness parameter then the cores evolve in line with the self-similar expectation, that they have cooled since $z=1$. \citet{McDonald2014} found that the temperature in the cluster core has increased relative to the characteristic cluster temperature, $T_{500}$, towards lower redshift. However, these clusters were split by the cuspiness parameter, with the ``cuspiest" $50$ per cent of clusters defined as CC. It is not clear how this division, rather than by a fixed threshold, changes the observed evolution.

\subsection{Pressure profiles}
\begin{figure*}
 \includegraphics[width=1.02\textwidth]{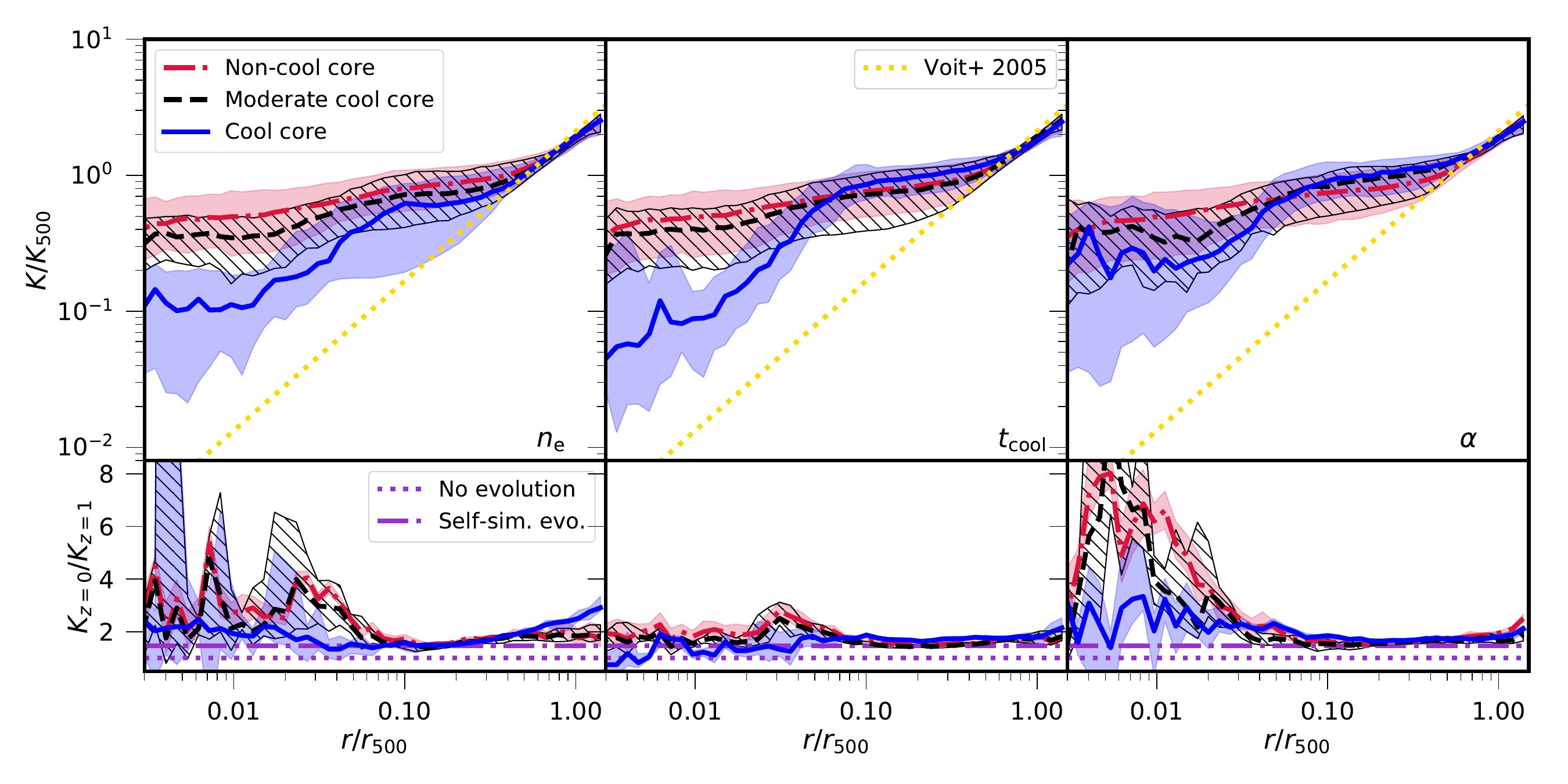}
 \caption{Median gas entropy profiles at $z=0$ (top row) and the ratio of the median entropy profiles at $z=0$ and $z=1$ (bottom row). The panels and line styles are the same as in Fig. \ref{fig:rho_profiles}. Additionally, we plot the non-radiative simulations of \citet{Voit2005} (yellow dotted) and find that all profiles asymptote to this result at $0.6r_{500}$. The entropy profiles between $z=1$ and $z=0$ increase by at least the self-similar expectation, rising beyond that in the core due to the temperature increase.}
 \label{fig:K_profiles}
\end{figure*}

In the top row of Fig. \ref{fig:P_profiles} we plot the median CC, MCC and NCC pressure profiles for the different criteria at $z=0$. A linear combination of density and temperature, for $r>0.2r_{500}$ the profiles are in good agreement for all criteria, as expected from the density and temperature profiles. Inside this radius the classification of CCs by the central electron number density yields a CC profile with a central pressure that is $30$ per cent larger than the MCC profile, a result of the increased central density. For the central cooling time and cuspiness parameter the decrease in central temperature offsets the modest increase in central density and all profiles have similar central pressures. The overall change in the normalization of the median CC profiles for the different criteria is due to the mass dependence of their scatter, which changes the median mass of the sample.

In addition, we plot the CC and NCC universal pressure profiles for the \textit{Planck} ESZ sample \citep{Planck_IR_V}. CCs were defined as having a central electron number density $n_{\mathrm{e}}>4\times10^{-2}\,\mathrm{cm}^{-3}$ \citep{Planck_ER_XI}. We calculate the observed universal profile using the listed parameters in \citet{Planck_IR_V}. For each simulated CC, MCC and NCC profile for each criteria we calculate the characteristic pressure, $P_{500}$, using the median mass of the sample. We then calculated the expected $P(r/r_{500})$ by multiplying the universal pressure profile by $P_{500}$. For all criteria the simulated and observed NCC profiles show good agreement. A like-with-like comparison classifying clusters by the central electron number density results in observed and simulated CC profiles that are in good agreement. However, using a different criterion leads to a disagreement between the simulated profile and the universal profile observed for CCs defined by the central electron number density. Hence, the properties of the average CC changes based on the chosen CC criterion.

In the bottom row we plot the ratio of the median profiles at $z=0$ and $z=1$. The self-similar evolution is given by $1/E^{8/3}(z=1)=0.215$. All profiles for the central cooling time and cuspiness parameter criteria are consistent with the self-similar expectation throughout the cluster volume. Classifying clusters by the central electron number density yields profiles consistent with the self-similar expectation for $r>0.1r_{500}$. Inside this radius the average cluster profile evolves less than expected, with central values of $0.76$, $0.68$ and $0.58$ at $0.01r_{500}$ for the median CC, MCC and NCC profiles, respectively. This deviation is driven by the lack of evolution in both the temperature and density profiles.

\begin{figure*}
 \includegraphics[width=1.02\textwidth]{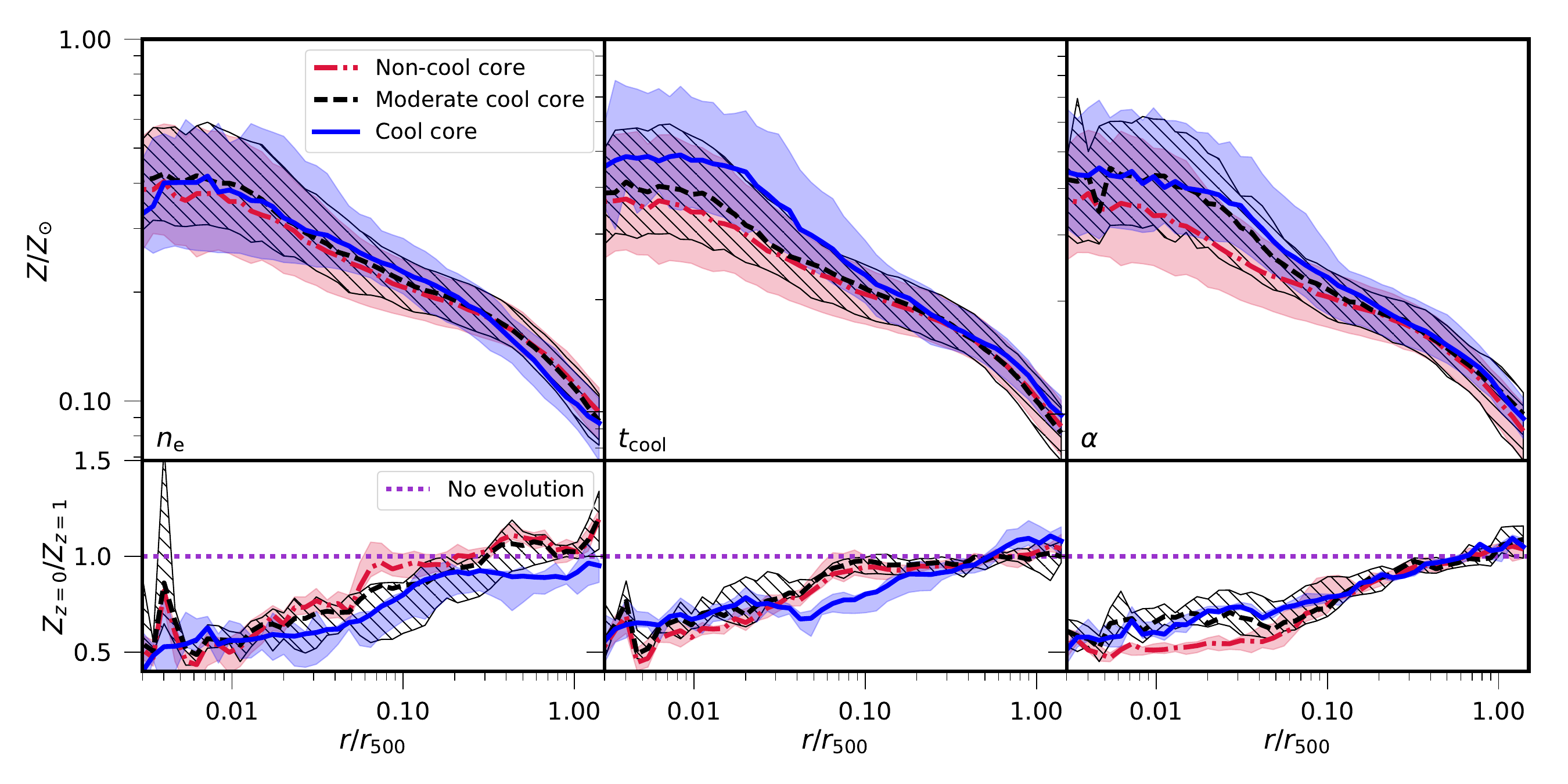}
 \caption{Median metallicity profiles at $z=0$ (top row) and the ratio of the median metallicity profiles at $z=0$ and $z=1$ (bottom row). We scale all abundances to the solar values of \citet{AndersGrevesse1989}. The line styles are the same as in Fig. \ref{fig:rho_profiles}. The lack of evolution for $r>0.15\,r_{500}$ is consistent with the idea of early enrichment.}
 \label{fig:Z_profiles}
\end{figure*}

\subsection{Entropy profiles}
We plot the median CC, MCC and NCC entropy profiles for the different criteria at $z=0$ in Fig. \ref{fig:K_profiles}. For $r>0.1r_{500}$ the profiles are in good agreement with each other for all criteria and asymptote to the non-radiative result at $0.6r_{500}$ \citep{Voit2005}. Inside $0.1r_{500}$ the increased density and decreased temperatures of the CC profiles yield a drop in the entropy profile compared to the NCC profiles. The extent of the drop depends on the defining criteria, with central cooling time CC profile a factor of $5$ lower than the NCC profile at $0.01r_{500}$, but the cuspiness parameter CC profile is only a factor $2$ lower.

The bottom panels show the ratio of the median profiles at $z=0$ and $z=1$ for the different CC criteria. The expected self-similar entropy evolution is given by $1/E^{-2/3}(z=1)=1.468$. We find that all profiles produce ratios that are consistent with the self-similar expectation when classifying clusters by their central cooling time, and greater than self-similar in the core of clusters defined by the cuspiness parameter. The entropy inside $r=0.1r_{500}$ increases faster than than expected because the clusters increase in temperature between $z=1$ and $z=0$.

\subsection{Metallicity profiles}
Finally, we examine the median metallicity profiles for the different criteria in the top row of Fig. \ref{fig:Z_profiles}. The profiles are normalized to the solar abundances of \citet{AndersGrevesse1989}. Although examined in \citet{Vogelsberger2017}, here we examine how the metallicity evolution changes for different CC criteria. Classifying clusters by their central cooling time we find evidence of a core in the CC profile for $r<0.1r_{500}$ with a central metallicity that is $25$ per cent higher than the MCC and NCC median profiles. The CC and MCC profiles show evidence of a core when defined by the cuspiness criterion and have marginally higher central metallicities compared to the NCC profile. The profiles are in good agreement throughout the cluster volume when split by their central electron number density, with all of them showing some hints of a core, i.e. they flatten to a constant central metallicity value.

In the bottom row we plot the ratios of the median profiles at $z=0$ and $z=1$. All profiles show the metallicity inside $0.15r_{500}$ has decreased since $z=1$, a result consistent with previous numerical work \citep[e.g.][]{Martizzi2016} but inconsistent with the latest observations \citep[e.g.][]{McDonald2016,Mantz2017}. This may be due to differences in how the subgrid and real AGN shape the ICM and distribute metals or the result of systematic uncertainties in the observations. Outside the core the profiles are consistent with minimal evolution in the iron abundance and the idea of early enrichment. There are marginal changes in the size of the deficit radius when CCs are classified by different criteria.

In summary, we find that the thermodynamic profiles in the core of simulated clusters have evolved since $z=1$. The extent of this evolution depends on the chosen CC criteria, as the mass dependence of the criterion impacts the median profiles. For example, defining CC via the same criterion as the observed profile yields a good agreement between the simulations and the observations, but differing criteria with opposite mass trends results in differing profiles. The departure of the simulated core profile evolution from the self-similar expectation is in agreement with the observed cluster core evolution, even if the exact scale of the departure differs between the simulations and the observations. Part of this discrepancy, e.g. the reduction in central metallicity, provides insight into how subgrid and real AGN differ in their shaping of the cluster volume. It may also be an indication that additional physical processes, such as anisotropic thermal conduction, must be included to correctly capture the formation and evolution of the cores of clusters.

\section{Conclusions}
\label{sec:concs}
We have examined the fraction of clusters simulated with the IllustrisTNG model that host a CC. We focused on the TNG300 level-1 periodic volume, which has a side length of $302\,\mathrm{Mpc}$ and a mass resolution of $1.1\times10^{7}\,\mathrm{M}_{\astrosun}$ and $5.9\times10^{7}\,\mathrm{M}_{\astrosun}$ for the gas and dark matter components respectively. We selected all clusters with a mass $M_{500}>10^{13.75}\,\mathrm{M}_{\astrosun}$, which yielded a sample of $370$ $(77)$ clusters at $z=0$ $(z=1)$. We then examined the CC fraction for $6$ different criteria commonly used in the literature (Section \ref{sec:results}), the evolution of the CC fraction with redshift (Section \ref{sec:CCevo}) and how the cluster core evolved in comparison to the rest of the cluster volume (Section \ref{sec:profs}). Our main results are as follows:
\begin{itemize}
 \item The $z=0$ CC fraction for the complete sample is in good agreement with previous numerical work \citep{Rasia2015,Hahn2017}. Selecting a sample of high-mass clusters we find that the CC fraction at $z=0$ decreases for $3$ criteria (central cooling time, central entropy excess and the cuspiness parameter) and increases for $3$ criteria (central electron number density and the concentration parameter within physical and scaled apertures) relative to the complete sample. The simulated high-mass sample is lower than the observed CC fraction for $4$ criteria (Fig. \ref{fig:M500-CCcrit}), and in reasonable agreement for $2$ criteria (central electron number density and the scaled concentration parameter). When fit with a linear relation $5$ criteria are consistent with weak or no mass dependence, with the concentration parameter with scaled apertures the exception. The scatter about the best-fit relation is mass dependent for the central electron number density and the concentration parameters.
 \item The simulated gas fractions are lower than observed at $0.01r_{500}$ and increase more rapidly than observed, with the cumulative profile reaching $86$ per cent of the universal fraction at $0.6r_{500}$ compared to the observed fraction that reaches $83$ per cent at $r_{500}$. This difference is likely due to the AGN feedback being more violent than in reality, explaining why the simulated CC fractions are lower than observed and some of the differences between the simulated and observed the criteria distributions (Fig. \ref{fig:cc_crit_comp}).
 \item The simulated central entropy excess distributions are single peaked, but a significant number of the clusters have effectively zero central entropy excess. Previous idealized numerical work and observations of filamentary molecular gas around BCGs support the idea that cold gas precipitates out of the hot phase once the ratio of the cooling time to the free-fall time reaches a small enough value and maintains a minimum central entropy. The simulated central entropy excess distribution may suggest that modelling the formation of a cold phase is an important component in reproducing cluster cores. However, we note that \citet{Panagoulia2014} found pure power laws entropy profiles if only radii that are sufficiently well resolved are considered, making the exact central profile of CC clusters uncertain.
 \item The correlations between different CC criteria is mass dependent, with the correlation increasing for more massive systems. For example, the correlation between central electron number density and the cuspiness parameter improves from $0.31\pm0.06$ for the full sample to $0.75\pm0.10$ when only clusters with $M_{500}>2\times10^{14}\,\mathrm{M}_{\astrosun}$ are selected.
 \item Examining the CC fraction as a function of redshift (Figs. \ref{fig:CC_ne_zred}-\ref{fig:CC_alpha_zred}) we find that the high-mass CC fractions are in tentative agreement with the observed CC fractions. However, linear fits to the CC fraction as a function of redshift reveal that the redshift evolution of the CC fraction for the high-mass sample is at least $2\sigma$ steeper than the observed evolution for the central electron density, the central cooling time and the central entropy. The simulated redshift evolution of the CC fraction defined by the concentration parameter within physical apertures is consistent with the observed evolution, but the normalization is offset at all redshifts. Finally, the redshift evolution of the CC fraction defined by the cuspiness parameter is poorest fit by a linear relation and the redshift evolution of the simulated sample is consistent with both flat and the negative redshift evolution of the observations. The redshift evolution of the CC fraction for the complete sample is steeper than the high-mass sample for $4$ of the $5$ criteria.
 \item We found no evidence that the fraction of CCs defined as relaxed is greater than the fraction of NCCs defined as relaxed (Fig. \ref{fig:Relaxation}). Defining relaxed clusters by the ratio of their kinetic to thermal energy, all samples yield a relaxed fraction that decreases with redshift, consistent with the expectation that the merger rate increases. The result holds for other theoretical definitions of a relaxed cluster. This result seems to be at odds with the idea that mergers solely drive the CC/NCC bi-modality and the observation that radio haloes only occur in NCC clusters.
 \item Comparing the pressure profiles to the observed universal pressure profile demonstrates that a like-with-like comparison is required. The differing mass dependence of the CC fractions leads to different central pressure profiles and either agreement or disagreement with the observed profile. A like-with-like comparison produces a good agreement between the simulated and observed CC pressure profiles.
 \item The thermodynamic profiles in the cores of simulated clusters have evolved to some extent between $z=1$ and $z=0$ (Figs. \ref{fig:rho_profiles}-\ref{fig:Z_profiles}), with the extent of the evolution depending on the chosen CC criterion. The simulated core evolution departs from the self-similar expectation for many profiles, with the direction of the departure in agreement with recent observations of the evolution of cluster cores \citep{McDonald2017}. These results indicate that the heating and radiative losses in the centre of the simulated clusters are not in balance in the simulation and point to differences in the way the subgrid AGN and real AGN shape the cluster volume.
\end{itemize}
We conclude that the IllustrisTNG model matches the observed CC fraction between $0.25<z<1.0$, but converts CCs to NCCs too rapidly compared to the observations. This results in it overpredicting the fraction of CCs at $z>1$ and underpredicting the CC fraction at $z<0.25$. In future work we will investigate the mechanisms responsible for converting a CC to a NCC as we found that the fraction of CCs and NCCs defined as relaxed were similar, suggesting mergers may not be solely responsible for the dichotomy and in tension with previous numerical work \citep[e.g.][]{Hahn2017}. It seems that subgrid models must continue to develop, especially for the treatment of AGN, and continue to include additional physical processes since essentially all numerical simulations, including IllustrisTNG, do not capture the evolution and thermodynamic profiles of clusters correctly. These physical processes include the impact of cosmic-rays \citep{Pfrommer2017}, anisotropic thermal conduction \citep[Barnes et al in prep.]{Kannan2016,Kannan2017}, outflows due to radiation pressure from the AGN \citep{Costa2017a,Costa2017b}, the formation of dust \citep[][Vogelsberger et al. in prep.]{McKinnon2016} and, as simulations push to higher resolution, and most importantly more accurate modelling of the interaction between the jet and the intracluster medium \citep{English2016,Weinberger2017b}.

\section*{Acknowledgements}
We thank Adam Mantz for making data available to us. DJB acknowledges support from STFC through grant ST/L000768/1. MV acknowledges support through an MIT RSC award, the support of the Alfred P. Sloan Foundation, and support by NASA ATP grant NNX17AG29G. RK acknowledges support from NASA through Einstein
Postdoctoral Fellowship grant number PF7-180163 awarded by the Chandra X-ray Center, which is operated by the Smithsonian Astrophysical Observatory for NASA under contract NAS8-03060. Simulations were run on the Hydra and Draco supercomputers at the Max Planck Computing and Data Facility, on the HazelHen Cray XC40-system at the High Performance Computing Center Stuttgart as part of project GCS-ILLU of the Gauss Centre for Supercomputing (GCS), and on the joint MIT-Harvard computing cluster supported by MKI and FAS. VS, RW, and RP acknowledge support through the European Research Council under ERC-StG grant EXAGAL-308037 and would like to thank the Klaus Tschira Foundation. The Flatiron Institute is supported by the Simons Foundation. SG and PT acknowledge support from NASA through Hubble Fellowship grants HST-HF2-51341.001-A and HST-HF2-51384.001-A, respectively, awarded by the STScI, which is operated by the Association of Universities for Research in Astronomy, Inc., for NASA, under contract NAS5-26555. JPN acknowledges support of NSF AARF award AST-1402480. The flagship simulations of the IllustrisTNG project used in this work have been run on the HazelHen Cray XC40-system at the High Performance Computing Center Stuttgart as part of project GCS- ILLU of the Gauss Centre for Supercomputing (GCS). Ancillary and test runs of the project were also run on the Stampede supercomputer at TACC/XSEDE (allocation AST140063), at the Hydra and Draco supercomputers at the Max Planck Computing and Data Facility, and on the MIT/Harvard computing facilities supported by FAS and MIT MKI.

%%%%%%%%%%%%%%%%%%%% REFERENCES %%%%%%%%%%%%%%%%%%
\bibliographystyle{mnras}
\bibliography{main}

%%%%%%%%%%%%%%%%%%%% APPENDIX %%%%%%%%%%%%%%%%%%

\appendix
\section{Impact of resolution on cool-core criteria}
\label{app:resolution}
\begin{figure*}
 \includegraphics[width=0.49\textwidth,keepaspectratio=true]{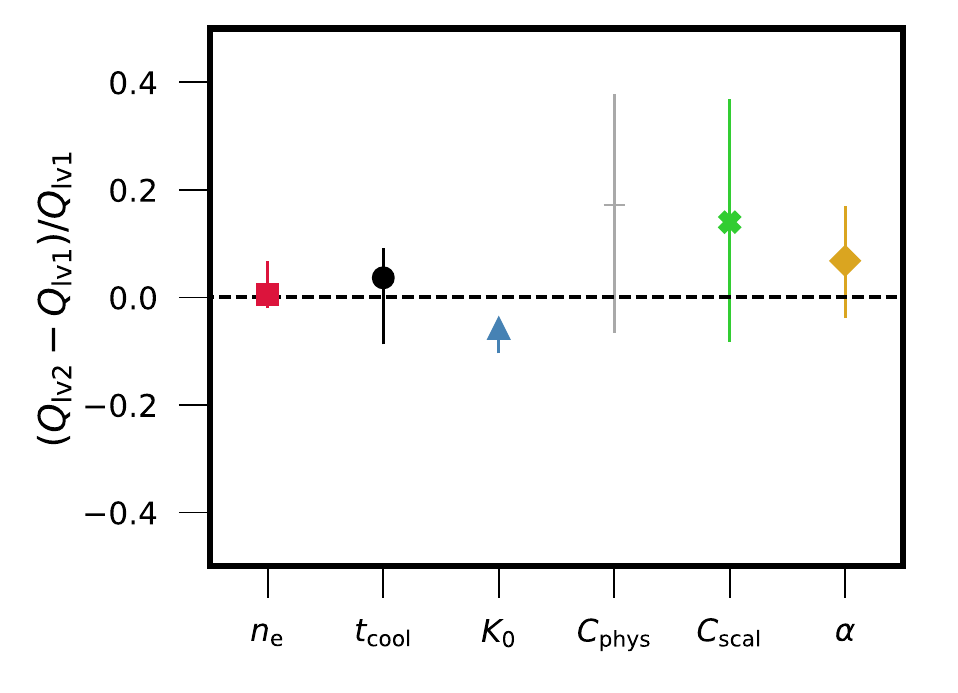}
 \includegraphics[width=0.49\textwidth,keepaspectratio=true]{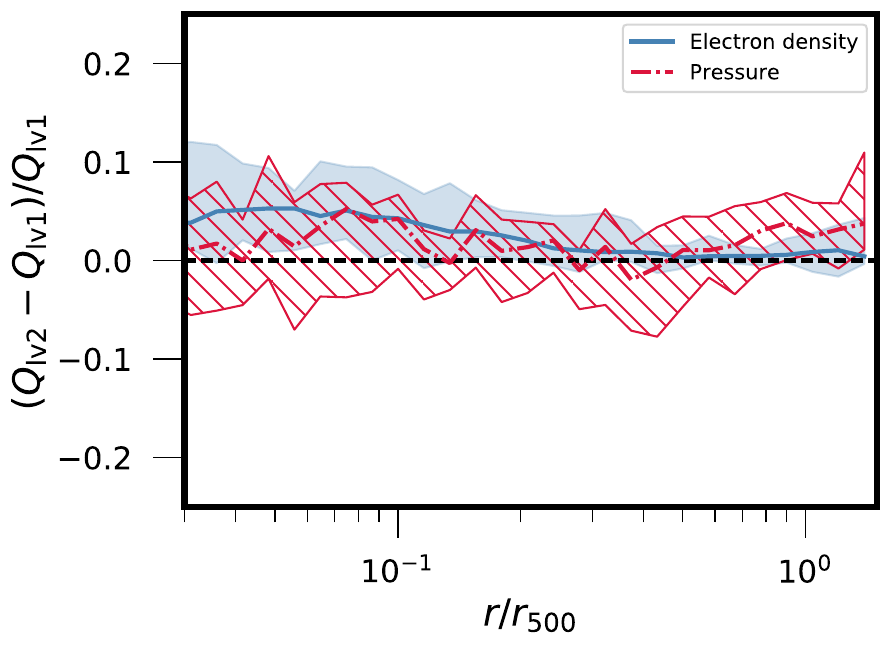} 
 \caption{Fractional difference in the median CC properties (left panel) and the median electron density (blue) and pressure (red) profiles (right panel) at $z=0$ between the level 1 and level 2 resolutions of the TNG300 simulation. The uncertainty is calculated by bootstrapping the samples $10,000$ times. With the exception of the central entropy excess the properties are consistent within the errors.}
 \label{fig:TNG300res}
\end{figure*}

To examine the impact of numerical resolution on the CC criteria distributions we examine the $z=0$ clusters in the TNG300 volume at the level 1 and level 2 resolutions. In the left panel of Fig. \ref{fig:TNG300res} we plot the fractional difference in the median value of the $6$ CC criteria used throughout this work. With the exception of the central entropy excess the median CC criteria values are consistent with each other. In the right panel we compare the median electron number density and pressure profiles from the level 1 and level 2 resolutions TNG300 volume. The pressure profiles are consistent with each other within the errors at all redshifts. The electron number density profiles are consistent for $r/r_{500}>0.1$. Inside of this radius we find that the high resolution simulation has a central electron number density that is $5$ per cent larger than the lower resolution simulation, but the uncertainty in the profile is of a similar magnitude. Therefore, we conclude that the results presented in this paper are relatively insensitive to numerical resolution.

\section{High-mass sample correlations and scatter}
\label{ap:cstab}
In Table \ref{tab:HMcorrelations} we plot the correlation coefficients and the scatter between the different CC criteria considered in this work for the high-mass sample, an analogue to Table \ref{tab:correlations} for the complete sample.

\renewcommand\arraystretch{1.1}
\begin{table*}
 \caption{Correlation and scatter about the best-fit power law for the different CC criteria at $z=0$ for the high-mass sample. The Spearman rank correlation coefficients, $r_{\mathrm{s}}$, are shown in the lower off-diagonal entries and the scatter, $\sigma_{\log_{10}}$, in the upper off-diagonal entries. Errors are computed by bootstrap resampling 10,000 times.}
 \begin{tabularx}{0.75\textwidth}{l R R R R R R}
 \hline
 Criterion & \multicolumn{1}{C}{$n_{\mathrm{e}}$} & \multicolumn{1}{C}{$t_{\mathrm{cool}}$} & \multicolumn{1}{C}{$K_{0}$} & \multicolumn{1}{C}{$C_{\mathrm{phys}}$} & \multicolumn{1}{C}{$C_{\mathrm{scal}}$} & \multicolumn{1}{C}{$\alpha$} \\
 \hline
 $n_{\mathrm{e}}$ & \multicolumn{1}{C}{\hspace{0.2cm}$-$} & $0.14\pm0.05$ & $0.22\pm0.05$ & $0.18\pm0.03$ & $0.29\pm0.09$ & $0.33\pm0.04$ \\
 $t_{\mathrm{cool}}$ & $0.98\pm0.01$ & \multicolumn{1}{C}{$-$} & $0.12\pm0.02$ & $0.36\pm0.16$ & $0.47\pm0.20$ & $0.33\pm0.06$ \\
 $K_{0}$ & $0.94\pm0.02$ & $0.97\pm0.02$ & \multicolumn{1}{C}{\hspace{0.2cm}$-$} & $0.40\pm0.18$ & $0.48\pm0.19$ & $0.27\pm0.05$ \\
 $C_{\mathrm{phys}}$ & $0.93\pm0.02$ & $0.94\pm0.02$ & $0.93\pm0.01$ & \multicolumn{1}{C}{\hspace{0.2cm}$-$} & $0.14\pm0.02$ & $0.25\pm0.03$ \\
 $C_{\mathrm{scal}}$ & $0.89\pm0.06$ & $0.85\pm0.08$ & $0.81\pm0.07$ & $0.91\pm0.03$ & \multicolumn{1}{C}{\hspace{0.2cm}$-$} & $0.20\pm0.02$ \\
 $\alpha$ & $0.75\pm0.10$ & $0.80\pm0.08$ & $0.84\pm0.06$ & $0.72\pm0.10$ & $0.54\pm0.15$ & \multicolumn{1}{C}{\hspace{0.2cm}$-$} \\
 \hline
 \end{tabularx}
 \label{tab:HMcorrelations}
\end{table*}
\renewcommand\arraystretch{1.0}

\section{Relaxed fractions and relaxation criteria}
\label{app:rlx}
\begin{figure*}
 \includegraphics[width=1.02\textwidth]{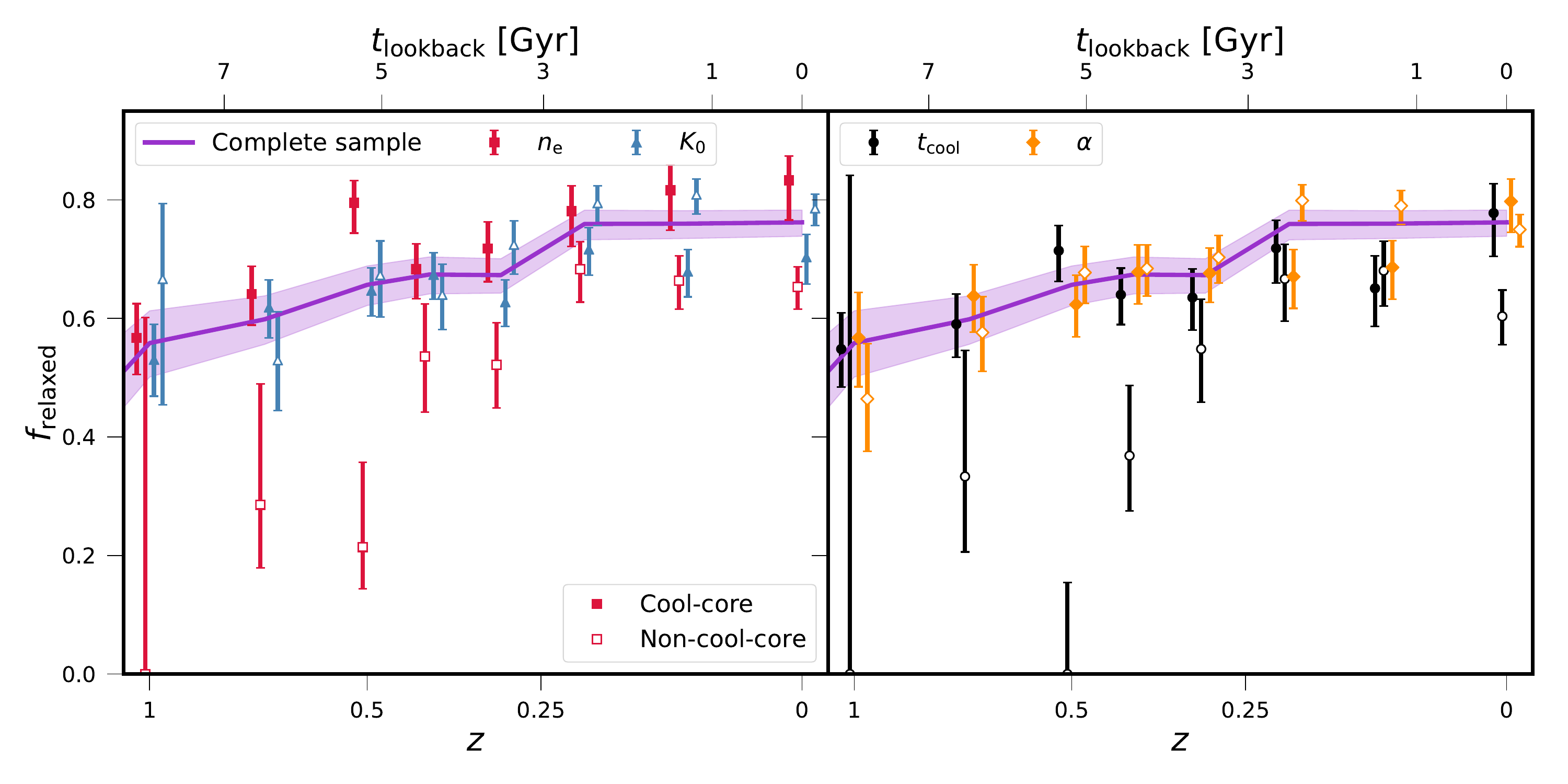}
 \caption{Fraction of clusters defined as relaxed via the combination of substructure mass fraction and the offset between the centre of mass and potential minimum as a function of redshift. The line and marker styles are the same as in Fig. \ref{fig:Relaxation}. The change of relaxation criteria does not impact the result that the complete sample, CC clusters and NCC clusters all have consistent relaxed fractions, regardless of CC definition.}
 \label{fig:Relaxation_FsXo}
\end{figure*}

In Fig. \ref{fig:Relaxation} we demonstrated that the fraction of CCs defined as relaxed was consistent with the fraction of NCC clusters and complete sample defined as relaxed. In Fig. \ref{fig:Relaxation_FsXo} we plot the fraction of clusters defined as relaxed by different theoretical criteria, the fraction of mass in substructures, $F_{\mathrm{sub}}$, and the offset between the centre of mass and the potential minimum scaled by the $r_{500}$, $X_{\mathrm{off}}$. Clusters are defined as relaxed if both $F_{\mathrm{sub}}<0.1$ and $X_{\mathrm{off}}<0.07$ \citep{Neto2007}. The overall fraction of clusters defined as relaxed has increased, however this can be reduced by making the relatively arbitrary threshold for relaxation stricter. In common with the relaxation criterion presented in Section \ref{sec:CCevo}, we find that the relaxed fraction decreases with redshift for all samples and, more importantly, that the fraction of CC clusters defined relaxed is consistent with the fraction of the complete sample and NCC clusters defined as relaxed, regardless of the chosen CC criteria.

\bsp    
\label{lastpage}
\end{document}